\documentclass[%
 reprint,
 amsmath,amssymb,
 aps,
 pra,
]{revtex4-1}

\usepackage{graphicx}
\usepackage{dcolumn}
\usepackage{bm}
\begin{document}


\title{The three-dimensional carrier-envelope-phase map of focused few-cycle pulsed Gaussian beams}


\author{Miguel A. Porras}
\affiliation{Grupo de Sistemas Complejos, ETSIME, Universidad Polit\'ecnica de Madrid, Rios Rosas 21, Madrid 28003, Spain}

\author{Zolt\'an L. Horv\'ath}
\affiliation{Department of Optics and Quantum Electronics, University of Szeged, D\'om t\'er 9., Szeged 6720, Hungary}

\author{Bal\'azs Major}
\affiliation{ELI-ALPS, ELI-HU Non-Profit Ltd., Dugonics ter 13, Szeged 6720, Hungary}
\affiliation{Department of Optics and Quantum Electronics, University of Szeged, D\'om t\'er 9., Szeged 6720, Hungary}
\email[]{Corresponding author: Balazs.Major@eli-alps.hu}


\date{\today}

\begin{abstract}
We derive an analytical expression that describes the complete three-dimensional carrier-envelope phase (CEP) distribution of in the focal volume of  ultrashort pulsed Gaussian beams focused by spherical mirrors or lenses. The focal CEP map depends on the so-called factor $g$ specifying the frequency-dependence of the beam width of the source few-cycle pulse, on its chirp and on the small chromatic aberration introduced by a lens without appreciably distorting or broadening the few-cycle pulse. We show how to tailor the CEP map of mirror-focused and lens-focused few-cycle pulses in order to produce negligible transversal and axial CEP variations in specific regions of the focal volume for phase-sensitive interactions of light with matter taking place in a volume or on a surface. We propose a quasi-achromatic doublet lens that can implement in practice these tailored CEP distributions.
\end{abstract}


\maketitle

\section{Introduction}
In this paper we determine the three-dimensional (3D) carrier-envelope-phase (CEP) map, or spatial variation of the CEP, of broadband, few-cycle pulses focused by mirrors or lenses, and describe several mechanisms to tailor it for specific applications. The CEP map is relevant to all nonlinear light-matter interaction experiments as high-harmonic or attosecond pulse generation, photoelectron emission from metal surfaces, nano-objects or structures; and in the control of chemical reactions in femtochemistry, where the CEP is a crucial parameter determining the outcome of the experiment \cite{Krausz2014NPhot,Baltuska2003Nature, Haworth2007NPhys, Ishii2014NC, Kruger2011Nature}. In gas high-order harmonic generation, for example, the XUV emission may even originate from a large portion of the focal volume, influencing its spectral content through phase matching
\cite{Ishii2014NC,Chini2014NPhot, Popmintchev2010NPhot, Rudawski2015EPJD}. Thus knowing and controlling the focal CEP variation, along with a consequent choice of the material sample position and extent, can be considered as relevant as controlling the CEP of the laser source.

It has happened slowly and with difficulty to stop identifying the variation of the CEP through a focus with Gouy's phase shift, probably because the former has its origin in the latter \cite{Tritschler2005OL, Yang2008COL}. Gouy's phase shift is an additional $-\pi$ phase shift of a monochromatic constituent through the focus that changes its phase velocity, while the CEP shift is the shift of the phase of the carrier oscillations at the time of pulse peak due to the different and space-dependent phase and group velocities \cite{Major2015JO}. The precise on-axis and off-axis CEP map of ideally focused few-cycle, transform-limited pulsed Gaussian beams was described in \cite{Porras2009OL}. The deviation from Gouy's phase shift is governed by the so-called factor $g$ of the input beam, also called Porras factor in \cite{Hoff2017NPhys}, which is determined by the variation of the beam width with frequency \cite{Porras2009OL}. On the experimental side, earlier measurements did not report conclusive deviations of CEP shift from Gouy's phase shift \cite{Lindner2004PRL, Tritschler2005OL, Wang2007OL}. Recent measurements based on spectral interferometry reported however strong deviations from Gouy's phase shift and their relation with the wavelength-dependent properties of the input beam \cite{Major2015AO}. Very recently, direct measurement of the on-axis and off-axis CEP map based on the CEP-sensitivity of the back-scattered photoelectrons from a nanotip scanning the focal volume \cite{Hoff2017NPhys} confirmed the strong deviations from the Gouy's phase shift and corroborated the predictions in \cite{Porras2009OL}, in particular, the predicted dependence with the $g$-factor. These observations underscore the need to properly characterize the laser source in use ---especially measuring its factor $g$--- to get advanced control on the CEP and hence on the allied CEP-sensitive light-matter interactions \cite{Hoff2017NPhys}. Unfortunately, there are no studies on how the factor $g$ determining the CEP map relates to the femtosecond laser source characteristics, so that no effective control can be exercised on the CEP under close-to-ideal focusing conditions as with spherical or parabolic mirrors.

Soon after the demonstration of CEP-sensitivity of different physical processes, methods have been proposed to attain some control on axial variations of the CEP based on dispersive material for specific applications \cite{Zapata-Rodriguez2008OE,Porras2009OE}.
We have previously studied the focusing of few-cycle pulses with lenses \cite{Porras2012APB,Porras2012JOSAB,MajorPhD} and determined the precise conditions for lenses to focus to transform-limited, few-cycle pulses traveling undistorted in the focal volume. We have also described how the  on-axis CEP variation differs from that using mirrors as an effect of the small chromatic aberration introduced by the lens \cite{Porras2012APB}. Although mirrors are thought to be generally preferable, chromatic focusing has recently been proposed as an additional control knob for high-order harmonic generation \cite{Holgado2017PRA}. Also, we have previously found that a small pulse chirp, lengthening the pulse by, e. g., less that one femtosecond, strongly affects the on-axis CEP variation \cite{Porras2012JOSAB}. Chirped femtosecond pulses has also been used to control high-order harmonic generation \cite{Salieres1998PRL, Chang1998PRA, Lee2001PRL, Mauritssom2004PRA, Holgado2016PRA}, chemical reactions \cite{Goswami2009CP} and  THz-generation \cite{Vidal2014JOSAB}. Chromatic aberration and chirp thus offer two control parameters, often used in femtosecond light-matter interaction experiments, to control the CEP map.

However, the capability of chromatic focusing and chirp to modify and control the CEP map has only been analyzed along the optical axis, and not off-axis, while nonlinear interactions generally take place in a volume, not just on-axis
\cite{Karimi2013OE, Krausz2014NPhot, Ishii2014NC,Hoff2017JOP}. It is thus of interest to these experiments to investigate how the complete 3D CEP map depends on $g$, chromatic aberration and chirp, so that specific axial and transversal CEP variations (e. g., no axial and/or radial CEP variation) required for particular CEP-sensitive interactions at specific target positions and extents can be tailored acting on these parameters.

Thus, in Sec. \ref{sec:3DCEPmap} we evaluate the complete 3D CEP map, recalling the conditions under which the few-cycle pulses can be focused without appreciable distortion. In Sec. \ref{sec:3DCEPmaptailor} we analyze the conditions under which the CEP map can be made flat, axially, transversally or both, in specific locations in the focal volume, with a focusing mirror or with a lens using small chirps and chromatic aberrations. In Sec. \ref{sec:doublet} we exemplify the results by showing these small pulse chirp and the tunable small chromatic aberration are easily provided by a separable achromatic doublet that would allow to generate these tailored CEP maps in practice.

\section{The three-dimensional CEP map}\label{sec:3DCEPmap}

If the beam to be focused is of sufficiently high quality, its spectrum can be assumed to be of the form of the collimated Gaussian beam $\hat E_{\rm in}(\omega,r)=\hat p_{\rm in}(\omega)\exp(-r^2/s_L^2)$, where $r$ is the radial coordinate from the optical axis, $\hat p_{\rm in}(\omega)$ is the on-axis spectrum, of bandwidth $\Delta\omega$ about the laser carrier frequency $\omega_0$. The beam spot size $s_L$ and the associated Rayleigh distance $L=\omega s_L^2/2c$ depend in general on frequency $\omega$. Since applications require careful focusing to a nearly diffraction-limited beam, we can assume that the spectrum after the focusing system is given, from well-known Gaussian beam formulas in the Debye approximation ($L\gg f$, or no appreciable focal shift), by $\hat E(\omega,r,z)=\hat p(\omega)(-f/q)\exp(i\omega r^2/2cq)\exp(i\omega z/c)$, where $z$ is the axial distance from the exit plane of focusing system, $q=Z-iL_R$ and $L_R=f^2/L$ are the complex beam parameter and Rayleigh distance of the focused beam, $f$ is the focal length, which may depend on frequency, $Z=z-f$ is an axial coordinate with origin at the focal point at each frequency, and $\hat p(\omega)$ may differ from $\hat p_{\rm in}(\omega)$ in the possible spectral changes introduced by the focusing system. The space-dependent parts of the spectral amplitude and phase at each point after the focusing system are then given by
\begin{eqnarray}
a(\omega,r,z)&=& \frac{f}{L_R}\frac{1}{[1+(Z/L_R)^2]^{1/2}}\exp[-r^2/s^2(Z)]\, , \\
\varphi(\omega,r,z)&=& -\frac{\pi}{2}- \tan^{-1}\left(\frac{Z}{L_R}\right) + \frac{\omega r^2}{2cR(Z)} + \frac{\omega}{c}z
\end{eqnarray}
where $s^2(Z)=s_f^2[1+(Z/L_R)^2]^{1/2}$, $s_f^2= 2cL_R/\omega$, $R(Z)=Z[1+ (L_R/Z)^2]$.

It has been shown \cite{Porras2012APB} that the above Gaussian-beam formulas describe accurately the spectrum of the focused pulsed beam if $s_L$ is small enough compared to the mirror or lens radius, for aperture effects and spherical aberration to be negligible. Under this condition, and if the focusing system is a lens, pulse broadening introduced by the lens material dispersion is substantially the same as that introduced by a slab of the lens material and thickness equal to the lens central thickness $D$, i. e., $\hat p(\omega) = \hat p_{\rm in}(\omega)\exp[i D n(\omega)]$, where $n(\omega)$ is the lens material refractive index \cite{Porras2012APB}. It is then possible to have a transform-limited, few-cycle pulse after the lens by simply pre-compensating for the dispersion introduced by that plate, as usually done with standard pulse-shaping techniques (e. g., GVD and/or TOD compensation, depending on the pulse duration and lens thickness $D$). We may also have a pulse with positive/negative chirp after the lens by under/over-compensating that material dispersion. It has also been shown \cite{Porras2012APB} that the distortions in the pulse shape about the focus originating from the lens chromatic aberration, measured by the parameter $\gamma= \omega (df/d\omega)/L_R$ (evaluated at the carrier frequency), are negligible if $|\gamma|\ll \omega_0/\Delta \omega$, a condition that is easily given with real lenses. Under these conditions, the only source of distortion in the pulse shape is the intrinsic distortion due to the transverse limitation of a focused beam, i. e., to diffraction, distortions that are generally very small, and have been shown to be accurately described by the first-order theory of diffraction effects in few-cycle pulses \cite{Porras2002PRE}. In this theory, the time-domain electric field, or inverse Fourier transform of $\hat E(\omega,r,z)$, is expressed as the enveloped carrier oscillations $E(t,r,z)=A(t,r,z)\exp\{-i[\omega_0 t-\varphi(\omega_0,r,z)]\}$, with an envelope given by
\begin{equation}\label{A}
A(t,r,z)\simeq a(\omega_0,r,z)p(\tau) + i a'(\omega_0,r,z)\frac{dp(\tau)}{d\tau}\, .
\end{equation}
Prime signs stand for derivative with respect to frequency, $p(\tau)$ is the inverse Fourier transform of $\hat p(\omega)$ evaluated at the local time $\tau= t-\varphi'(\omega_0,r,z)$. The second term in the rhs of Eq. (\ref{A}) accounts for the small diffraction-induced changes in the pulse shape about the focus.

The CEP at a point $(r,z)$ is the phase of the carrier oscillations at the time of pulse peak, i. e., $\Phi(r,z)=-\omega_0[\tau(r,z)+\varphi'(\omega_0,r,z)]+ \varphi(\omega_0,r,z)+ \phi(r,z)$, where $\tau(r,z)$ is the time at which $|A(\tau,r,z)|$ peaks and $\phi(r,z)$ the argument of $A(\tau,r,z)$ at that time. Thus, if we take the focal point at the carrier frequency ($r=0, z=f(\omega_0)\equiv f_0$) (focal point for short) as a reference point, the CEP shift from the focal point is given by $\Delta\Phi(r,z)=\Phi(r,z)-\Phi(0,f_0)$, which can be separated in the two contributions
\begin{eqnarray}\label{PHI1}
\Delta\Phi_1(r,z)&=& [-\omega_0\varphi'(\omega_0,r,z)+\varphi(\omega_0,r,z)]\nonumber \\ &-&[-\omega_0\varphi'(\omega_0,0,f_0)+\varphi(\omega_0,0,f_0)]\,,
\end{eqnarray}
and
\begin{eqnarray}\label{PHI2}
\Delta\Phi_2(r,z)&=& -\omega_0[\tau(r,z)-\tau(0,f_0)] + \phi(r,z) - \phi (0,f_0) \nonumber \\
 &\equiv& -\omega_0 \Delta\tau(r,z) + \Delta \phi(r,z)\, .
\end{eqnarray}
The first contribution is independent of the shape $p(\tau)$ of the pulse, the only one considered in \cite{Porras2009OL} for focusing without chromatic aberration, and has been verified experimentally to describe the actual CEP map with transform-limited (unchirped Gaussian) pulses \cite{Hoff2017NPhys} (which implies that the second contribution is negligible with transform-limited pulses). The effect of chromatic aberration on the CEP was considered in \cite{Porras2012APB}, but limited to on-axis points ($r=0$). The second contribution includes the effect of the small pulse reshaping, and depends on the pulse shape itself. For on-axis points only, it has been shown that any small amount of chirp in the focused pulse drastically alters the CEP map \cite{Porras2012JOSAB}. Here we evaluate the two contributions for off-axis points with chirped or unchirped Gaussian pulses, providing thus a complete description of the CEP map about the focus.

Evaluation of Eq. (\ref{PHI1}) is a straightforward but lengthy (and tedious) exercise of algebra and calculation of derivatives, of which we summarize the most relevant points. With the normalized variables $\zeta=Z/L_R$ and $\rho=r/s(Z)$ the spectral phase reads as $\varphi=-\pi/2-\tan^{-1}(\zeta) + \rho^2\zeta + (\omega/c)z$. Both  $\zeta$ and $\rho$ depend on $\omega$ because $f$, and $L$, and hence $L_R$, $Z=z-f$ and all other quantities in the Gaussian-beam formulas where they appear, depend on $\omega$. Careful evaluation of their derivatives yield  $\zeta'=-(f'/L_R)- \zeta (L'_R/L_R)$ and $(\rho^{2})^{\prime}=\rho^2[1/\omega-(L'_R/L_R)-2\zeta\zeta'/(1+\zeta^2)]$. After long algebra, the first contribution to the CEP shift from the focal point is obtained to be
\begin{eqnarray}\label{CEPSHIFT1}
\Delta\Phi_1(r,z) &=& -\tan^{-1}\left(\frac{Z}{L_R}\right) \nonumber \\
                       &+& \frac{1-2\frac{r^2}{s^2(Z)}}{1+\left(\frac{Z}{L_R}\right)^2} \left[g\left(\frac{Z}{L_R}\right)+ \gamma \left(\frac{Z}{L_R}\right)^2 \right] \nonumber \\
                       &+& \gamma \frac{r^2}{s^2(Z)} ,
\end{eqnarray}
where
\begin{equation} \label{eq:g_gamma}
g= -\frac{L'_R}{L_R}\omega\,, \quad \gamma= \frac{f'}{L_R}\omega \,,
\end{equation}
and where {\it all frequency-dependent quantities are evaluated at the carrier frequency $\omega_0$,} e. g., $Z=z-f_0$ is the axial distance from the focal point.

Similarly, the spectral amplitude, expressed in the dimensionless variables, is given by $a= (f/L_R) \exp(-\rho^2)/(1+\zeta^2)^{1/2}$. After some algebra, its derivative with respect to frequency can be written as $a'=h(\rho,\zeta)a$, where
\begin{equation}
h(\rho,\zeta)=h(\zeta)-\rho^2\left(\frac{1}{\omega}- \frac{L'_R}{L_R}- \frac{2\zeta\zeta'}{1+\zeta^2}\right)\, ,
\end{equation}
\begin{equation}
h(\zeta)= \frac{f'}{f}\left(1+ \frac{f}{L_R}\frac{\zeta}{1+\zeta^2}\right) - \frac{L'_R}{L_R}\left(1- \frac{\zeta^2}{1+\zeta^2}\right) \, ,
\end{equation}
and where all frequency-dependent quantities are evaluated at the carrier frequency.
The complex envelope in Eq. (\ref{A}) then yields $A\simeq\left[p(\tau)+ i h(\rho,\zeta)dp/d\tau\right]a$, which, being a first-order approximation (second order derivatives are neglected), is conveniently replaced with $A \simeq p\left[\tau+ih(\rho,\zeta)\right]a$, having the same first-order approximation. Assuming that the on-axis pulse immediately after the lens is the Gaussian pulse $p(\tau)=(\Delta T/b)\exp(-\tau^2/b^2)$, where $b^2=\Delta T^2 -2iC$, $C$ is a residual chirp, and $\Delta T$ is the transform-limited duration (actual duration $\Delta T_C =\Delta T[1 + (2C/\Delta T^2)^2]^{1/2}$), the reshaped real amplitude's temporal shape in the focal region is found to be
\begin{equation}
|A|\simeq \frac{\Delta T}{b}\exp\left[\frac{h^2(\rho,\zeta)}{\Delta T^2}\right]\exp\left\{-\frac{\left[\tau- \left(\frac{2C}{\Delta T^2}\right)h(\rho,\zeta)\right]^2}{\Delta T_C^2}\right\}.
\end{equation}
Thus, reshaping due to the strong localization in the focal region consists of a drift of the time of pulse peak from point to point of space given by $\tau(\rho,\zeta)= (2C/\Delta T^2)h(\rho,\zeta)$. This expression allows to evaluate $-\omega_0[\tau(r,z)-\tau(0,f_0)]= -\omega_0(2C/\Delta T^2)[h(\rho,\zeta)-h(0,0)]$ in Eq. (\ref{PHI2}) for the second contribution to the CEP shift from the focal point. Similar calculations allow to conclude that the difference of phases of the envelopes at the drifted peaks, $\phi(r,z)-\phi(0,f_0)$, does not give a significant contribution to the CEP shift. After some algebra, Eq. (\ref{PHI2}) for the second contribution to the CEP shift then yields
\begin{eqnarray}\label{CEPSHIFT2}
\Delta\Phi_2(r,z) &=& \frac{2C}{\Delta T^2}\frac{1-2\frac{r^2}{s^2(Z)}}{1+\left(\frac{Z}{L_R}\right)^2} \left[-\gamma\left(\frac{Z}{L_R}\right)+ g \left(\frac{Z}{L_R}\right)^2 \right] \nonumber \\
                       &+& \frac{2C}{\Delta T^2} (1+g) \frac{r^2}{s^2(Z)}\,,
\end{eqnarray}
where again all quantities are evaluated at the carrier frequency. The total CEP shift form the focal point is finally obtained to be
\begin{eqnarray}\label{CEPSHIFT}
\Delta\Phi(r,z) &=& -\tan^{-1}\left(\frac{Z}{L_R}\right) \nonumber \\
                       &+& \frac{1-2\frac{r^2}{s^2(Z)}}{1+\left(\frac{Z}{L_R}\right)^2} \left[G\left(\frac{Z}{L_R}\right)+ \Gamma \left(\frac{Z}{L_R}\right)^2 \right] \nonumber \\
                       &+& \left[\Gamma + \frac{2C}{\Delta T^2}\right] \frac{r^2}{s^2(Z)} ,
\end{eqnarray}
where
\begin{equation}\label{GGAMMA}
G\equiv g-\gamma\frac{2C}{\Delta T^2}\,, \quad \Gamma\equiv \gamma+ g\frac{2C}{\Delta T^2}\,.
\end{equation}
These equations provide the complete 3D CEP map for the focusing conditions of interest in many applications to a nearly diffraction-limited and transform-limited pulsed beam. The most relevant novelty compared to the on-axis formula is the term in the third row that makes the CEP at the focal plane to present, in general, a quadratic variation with radial distance. Out of the focal plane, the CEP has also a positive or negative quadratic variation, depending on the specific value of $Z$. Along caustic surfaces $r/s(z)=\mbox{const}$, the CEP variation is more pronounced or less pronounced than Gouy's phase, depending on the value of $g$ of the input beam and the specific caustic surface, and as described elsewhere \cite{Porras2009OL,Porras2012JOSAB,Hoff2017NPhys}. The CEP shift along the caustic surface $r/s(Z)=1/\sqrt{2}$ always equals to Gouy's phase shift. For further development, we rewrite (\ref{CEPSHIFT}) more compactly as
\begin{equation}\label{CEPSHIFTC}
\Delta \Phi = -\tan^{-1}\zeta + \frac{\zeta G + \zeta^2 \Gamma}{1+\zeta^2}(1- 2\rho^2) + P \rho^2,
\end{equation}
where
\begin{equation}\label{P}
P=\Gamma + 2C/\Delta T^2=\gamma + (g+1)\frac{2C}{\Delta T^2}.
\end{equation}

\section{Tailoring the three-dimensional CEP map}\label{sec:3DCEPmaptailor}

The CEP map can be used and manipulated in a number of ways, depending on the particular application. For example, it is generally desirable to have a constant CEP in the volume of the target to avoid CEP-integration effects that may wash out the sensitivity to the CEP of the light-matter interaction. One may then evaluate the CEP standard deviation in the target volume and minimize it. In the following we assume a laser source of certain factor $g$ (that cannot easily be modified), a target at position $Z$ of narrow axial thickness compared to $L_R$, and large transversal size compared to $s(Z)$. We can also consider an light-matter interaction that depends on a certain power $n$ of the intensity $I^n\propto \exp[-2n r^2/s^2(Z)]= \exp(-2n\rho^2)$. This is an alternative way to consider a transversally limited target of effective radius $s(Z)/\sqrt{n}$ smaller than the beam radius, e. g., negligible transversal size for $n\rightarrow \infty$. The effective CEP in the thin sample, or average CEP with $I^n$ is found to be
\begin{equation}\label{EFFCEP}
\langle\Delta\Phi\rangle = -\tan^{-1}\zeta + \frac{\zeta G + \zeta^2 \Gamma}{1+\zeta^2}\left(1- \frac{1}{n}\right) + \frac{P}{2n}\,,
\end{equation}
with a transversal standard deviation given by
\begin{equation}\label{SV}
\langle (\Delta\Phi - \langle\Delta\Phi\rangle)^2\rangle^{1/2} =  \frac{1}{n}\left|\frac{G\zeta+\Gamma \zeta^2}{1+\zeta^2}- \frac{P}{2} \right|.
\end{equation}
For a given sample position $\zeta$, we may wish to have no transversal variation of the CEP, no local axial variation of the CEP, or both, acting on the two control parameters at hand, namely, pulse chirp and lens chromatic aberration. No transversal variation requires
\begin{equation}\label{NOR}
2\zeta G + 2\zeta^2\Gamma - (1+\zeta^2)P=0\,,
\end{equation}
locally vanishing axial variation of the effective CEP at position $\zeta$,  and in particular vanishing on-axis CEP variation for $n\rightarrow\infty$, requires
\begin{equation}\label{NOA}
-(1+\zeta^2)+ (1-\zeta^2)G\left(1-\frac{1}{n}\right) + 2\zeta\Gamma\left(1- \frac{1}{n}\right)=0\,,
\end{equation}
as obtained by equating to zero the derivative of Eq. (\ref{EFFCEP}) with respect to $\zeta$.

\subsection{Design of the CEP map with focusing mirrors}

Since in most of experiments mirrors are used to focus few-cycle pulses, we first analyze if the above conditions can be given with a mirror, in which case $\gamma=0$. Then $G=g$, $\Gamma = g(2C/\Delta T^2)$ and $P=(1+g)(2C/\Delta T^2)$. From Eq. (\ref{NOA}), no axial CEP variation at the focus $\zeta=0$ occurs, irrespective of the chirp, only with specific value $g=1/(1-1/n)$ of the input pulse, or $g=1$ for the on-axis CEP, and therefore it is not generally possible. From Eq. (\ref{NOR}), the CEP across the focal plane is constant only if $P=(g+1)(2C/\Delta T^2) = 0$, i. e., requires a transform-limited pulse. Any chirp induces a CEP variation across the focal plane. Axially and radially constant CEP occurs only given with input laser pulses with $g=1$ and $C=0$.

\begin{figure}[t]
\includegraphics[width=8cm]{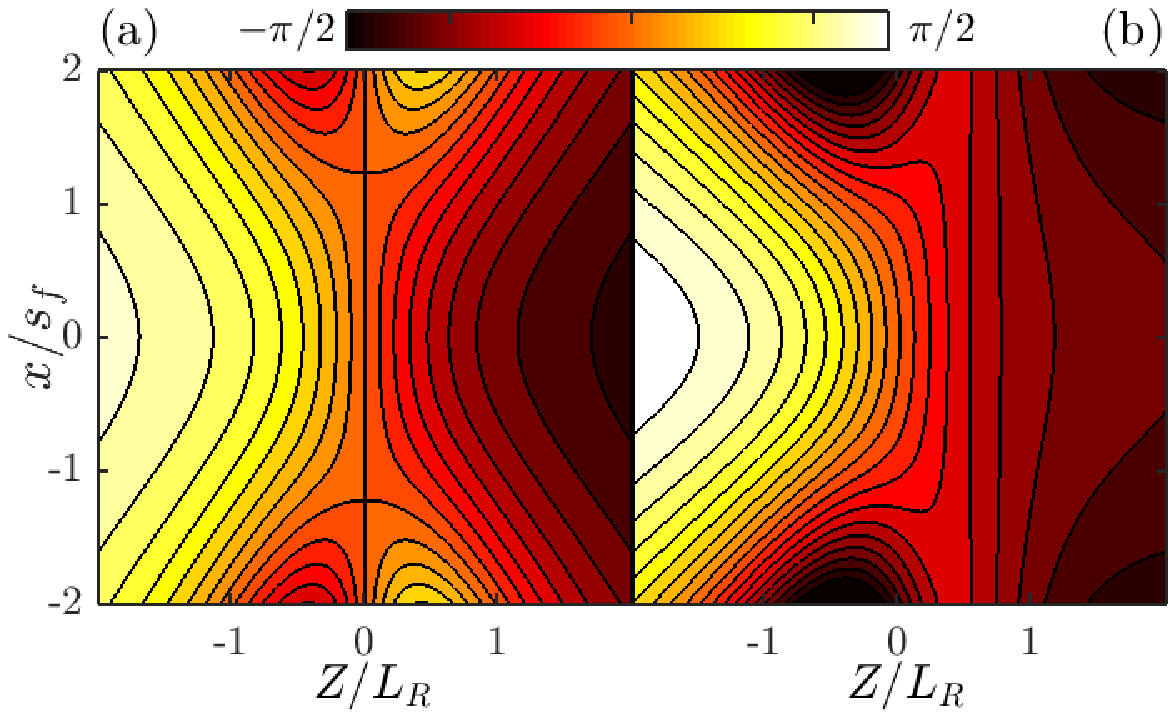}\\
\includegraphics*[width=4.3cm]{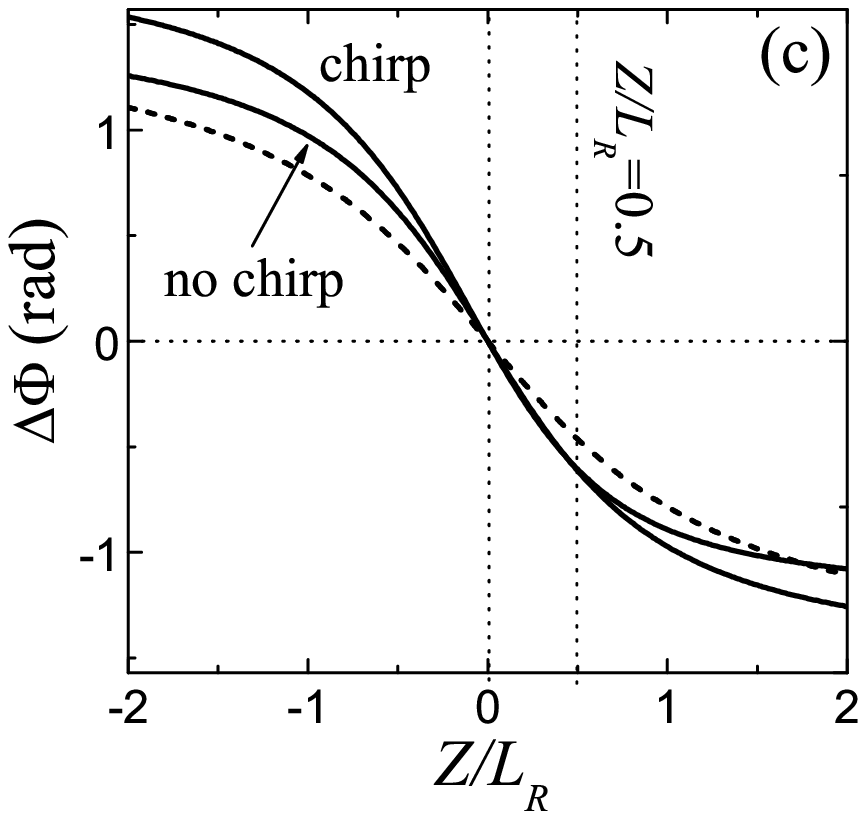}\includegraphics*[width=4.3cm]{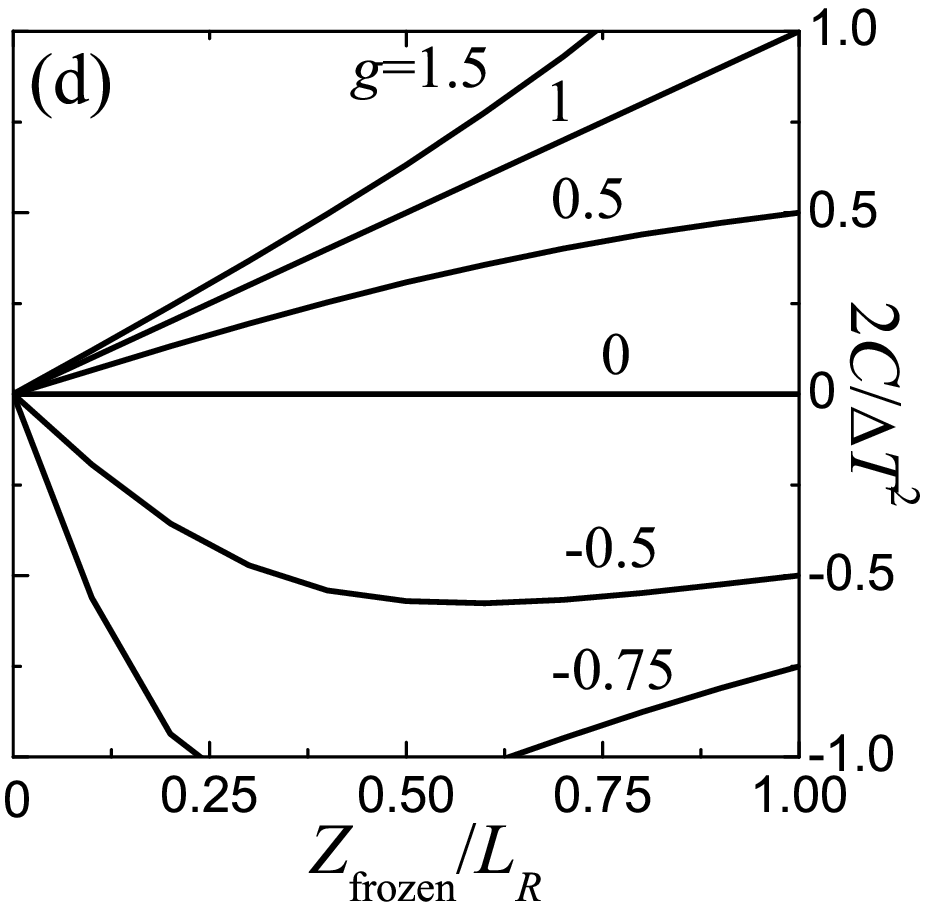}
\caption{\label{Fig1} For spherical mirror focusing, focal CEP maps (cut along $y=0$ plane) for input beam with $g=-0.5$, (a) transform-limited pulse, $2C/\Delta T^2=0$, and (b) chirped pulse, $2C/\Delta T^2=-0.571$. The CEP is transversally flat at the focal plane in (a) and at $Z=L_R/2$ in (b). (c) Axial variation of the CEP in the two above cases (solid curves) compared to Gouy's phase shift (dashed curve). (d) Needed relative chirp as a function of the position $Z_{\rm frozen}$ at which the CEP is constant transversally, for a few values of $g$ of the input pulse.}
\end{figure}

In many experiments the target (e. g., a gas nozzle) is placed slightly out of focus, preferably in its second half. Equation (\ref{NOR}) for transversally flat CEP at $0<|\zeta|<1$ in the case of a mirror yields the needed relative pulse chirp
\begin{equation}\label{NORM}
\frac{2C}{\Delta T^2} = \frac{2\zeta g}{g(1-\zeta^2)+(1+\zeta^2)}\,.
\end{equation}
Figure \ref{Fig1}(a) and (b) are examples of CEP maps in the focal volume of input focused pulses with $g=-0.5$, having radially flat CEP at $\zeta=0$ with transform-limited pulses ($C=0$), and radially flat CEP at the middle of the second half of the focal region with slightly chirped pulses ($2C/\Delta T^2 =-0.571$). Figure \ref{Fig1}(c) shows the corresponding on-axis axial variations, and FIG. \ref{Fig1}(d) the relative chirp $2C/\Delta T^2$ to have transversally flat CEP at different positions $\zeta_{\rm frozen}$ in the second half of the focal region for a few (supposedly) typical values of the factor $g$ of the input beam. The relative chirp providing radially flat CEP in the first half of the focal region has opposite sign. As seen, chirping the pulse could be useful in practice (does not entail important pulse broadening) to have transversally flat CEP at a desired axial position in the focal region for input pulses with $g>-1$.

\begin{figure}[t!]
\includegraphics[width=4.3cm]{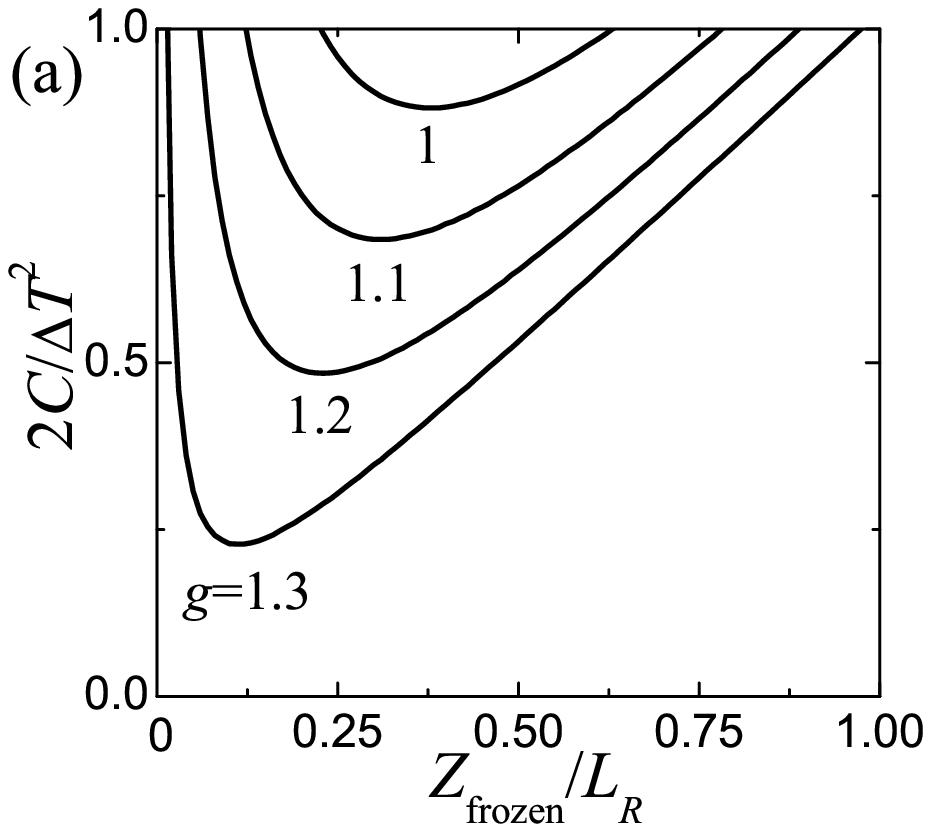}\includegraphics*[width=4.4cm]{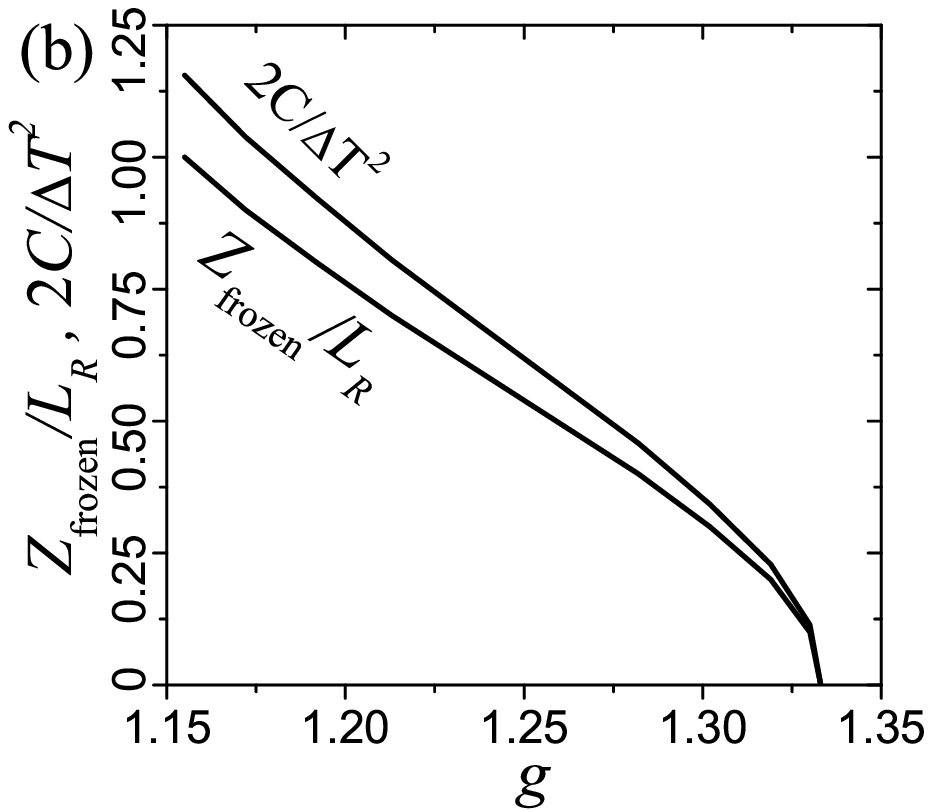}
\includegraphics[width=4.1cm]{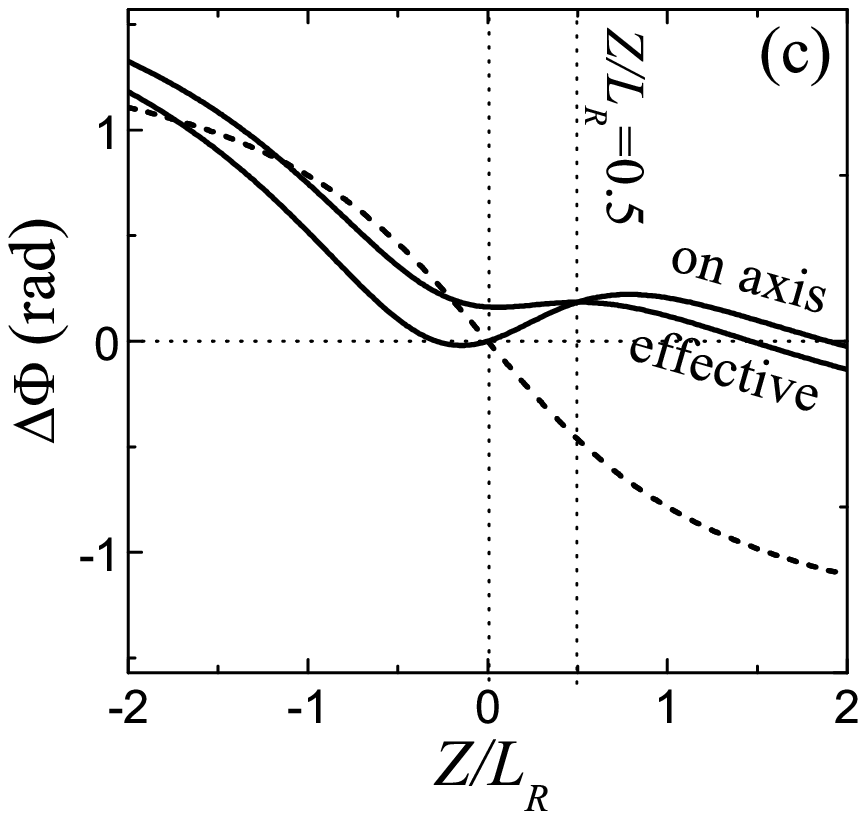}\includegraphics*[width=4.4cm]{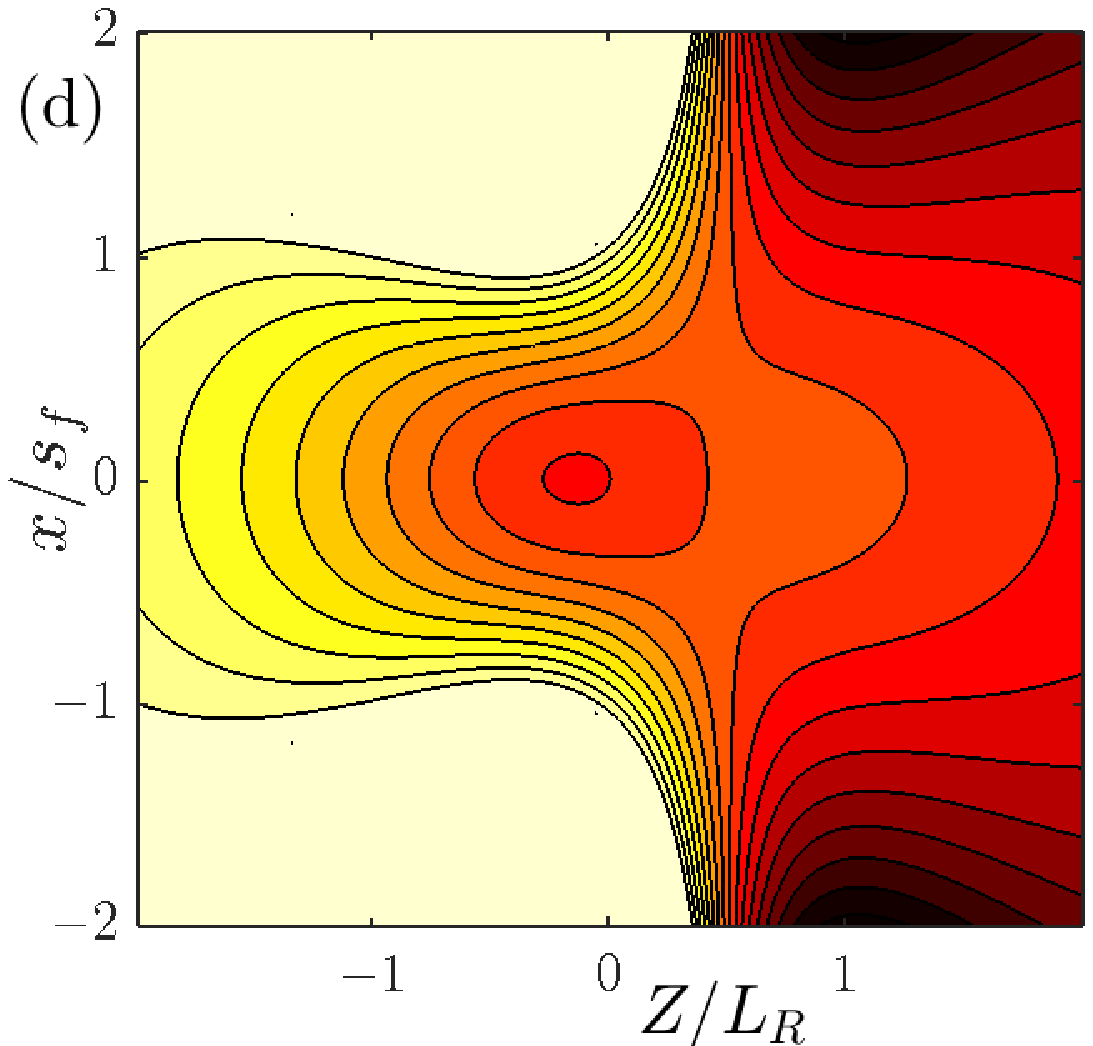}
\caption{\label{Fig2} (a) Needed relative chirp as a function of the position $Z_{\rm frozen}$ at which the effective CEP with $n=4$ is locally constant axially, for a few values of $g$ of the input beam for which this chirp is relatively low. (b) Needed relative chirp to freeze the effective CEP with $n=4$ both axially and transversally, and the position within the focus at which this happens, as functions of the values of $g$ for which this effect is possible. (c) On-axis and effective CEP with $n=4$ (solid curves) for $g=1.26$ and $2C/\Delta T^2=0.574$ and Gouy's phase (dashed curve), showing locally constant effective CEP at $Z=L_R/2$. (d) CEP map in the focal region for $g=1.26$ and $2C/\Delta T^2=0.574$, showing also transversally constant CEP at $Z=L_R/2$.}
\end{figure}

We instead may wish to have a CEP with no axial variation at a certain position $0<|\zeta|<1$ in the focal volume. Condition (\ref{NOA}) yields the chirp
\begin{equation}\label{NOAM}
\frac{2C}{\Delta T^2}=\frac{1+\zeta^2- (1-\zeta^2)g\left(1-\frac{1}{n}\right)}{2\zeta g \left(1-\frac{1}{n}\right)}\,.
\end{equation}
Unfortunately, the chirp values are considerably large for most of values of $g$. Only for $g>1$, reasonably small chirp produces locally constant CEP within the focal region, as shown in FIG. \ref{Fig2}(a) for the effective CEP with $n=4$ and in the second half of the focal region with positive chirps (opposite chirps produce the same effect in the first half).

Of particular interest is the situation in which the CEP is constant transversally and axially. Equating (\ref{NORM}) and (\ref{NOAM}), this effect is seen to be possible in the focal region only with an input beam with $g\ge 1.15$ at a certain position $\zeta$ in $[-1,1]$ (where the curves in FIG. \ref{Fig1}(d) and FIG. \ref{Fig2}(a) intersect) that depends on the particular value of $g$. This relevant position in the second half of the focus and the required chirp are depicted in FIG. \ref{Fig2}(b) as functions of $g$. The curves ends at $g=1/(1-1/n)$ at focus with no chirp. For example, with an input pulse with $g=1.26$, the chirp $2C/\Delta T^2=0.574$ leads to the axially frozen effective CEP with $n=4$ at $\zeta=0.5$, as seen in FIG. \ref{Fig2}(c). The transversally and axially flat CEP map at $\zeta=0.5$ is seen in FIG. \ref{Fig2}(d).

These results underscore the need to characterize the femtosecond laser source in use by measuring its factor $g$. By doing this we will not only know the possible CEP maps about the focus of the mirror, but will also know if the CEP map can be adapted to particular applications using small amounts of chirp as a control knob, or suitably positioning the sample according to the CEP maps.

\subsection{Design of the CEP map with lenses}

\begin{figure}[b!]
\includegraphics*[width=4.3cm]{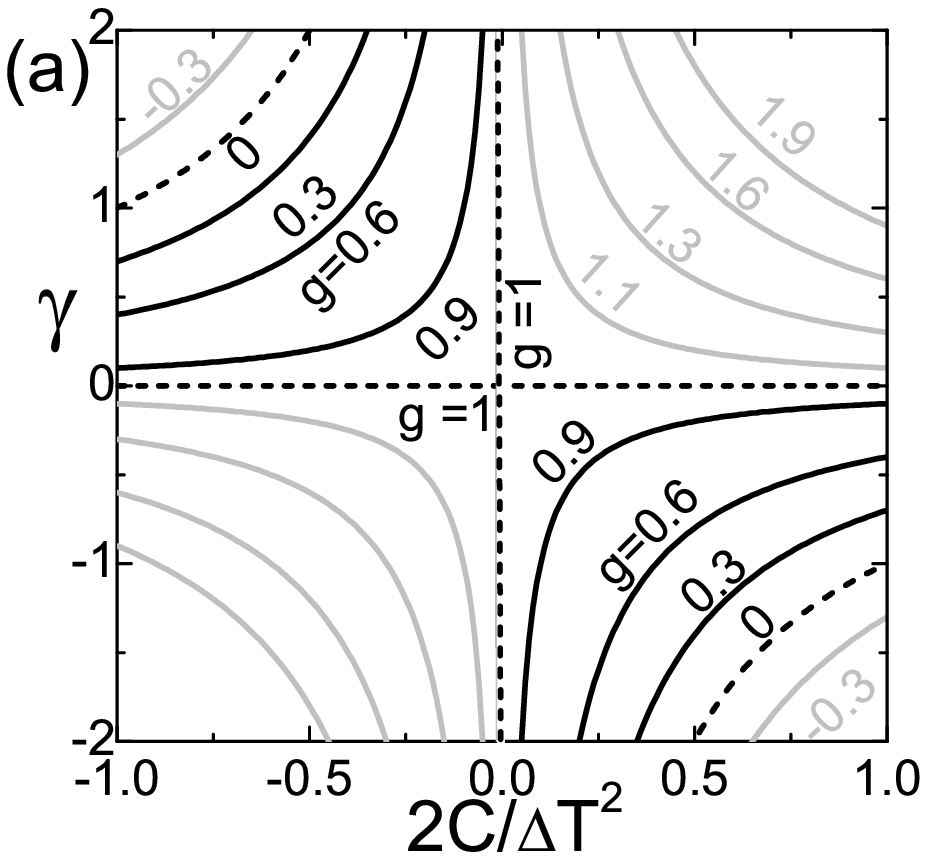}\includegraphics*[width=4.3cm]{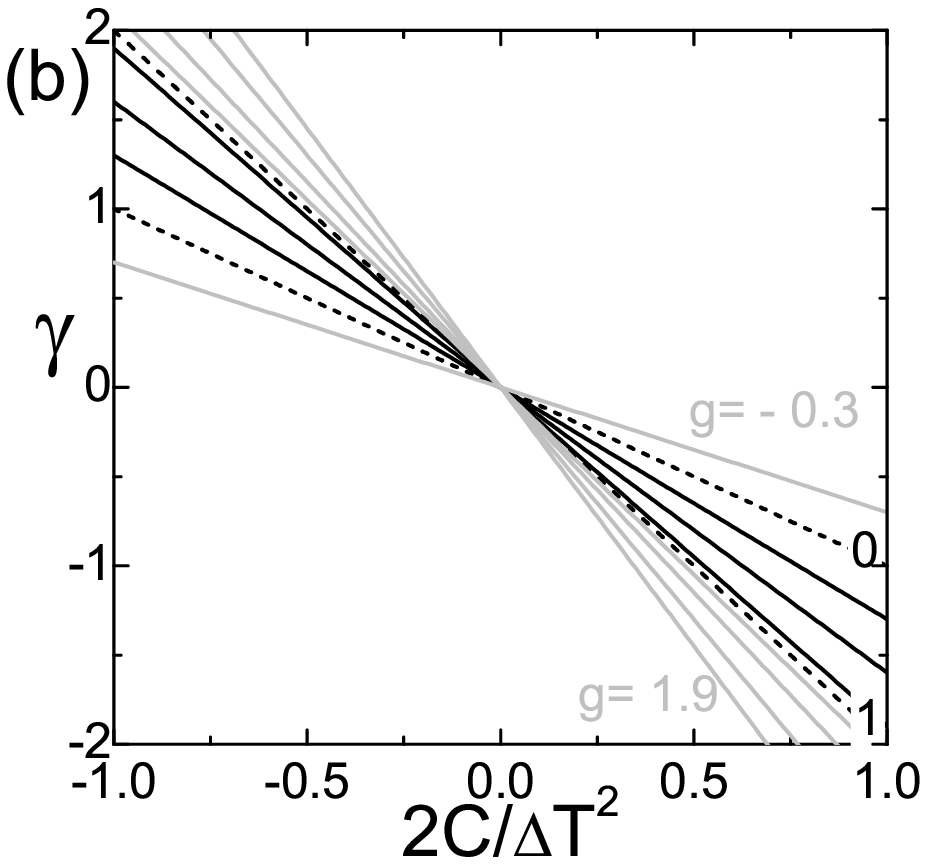}
\includegraphics*[width=4.3cm]{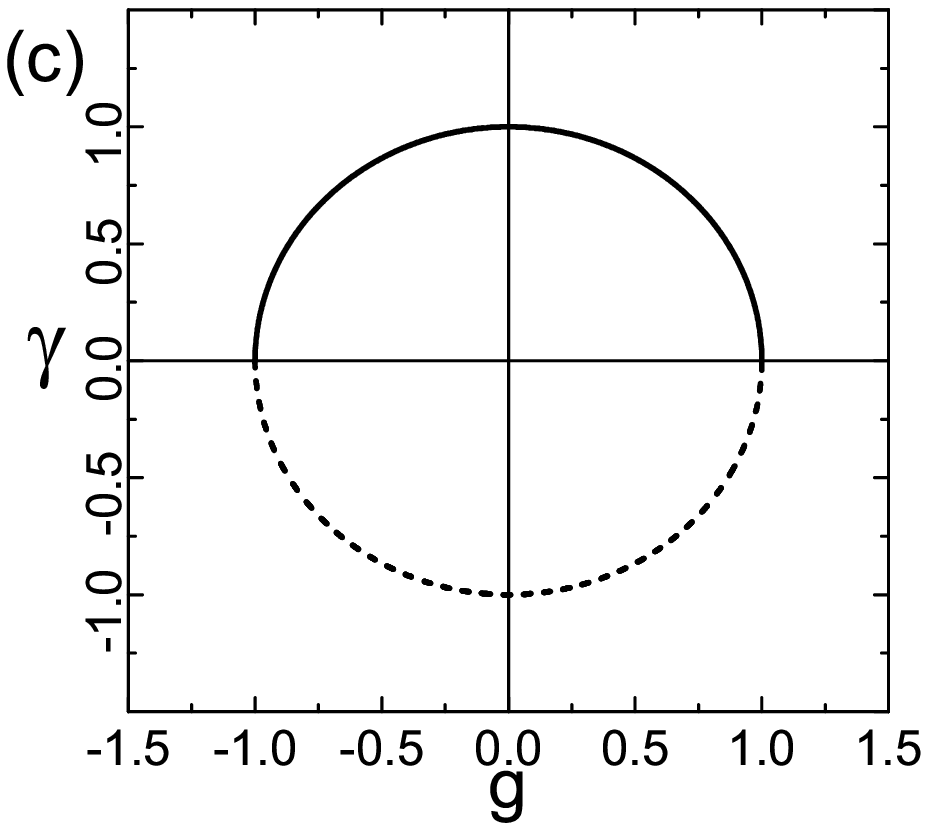}\includegraphics*[width=4.3cm]{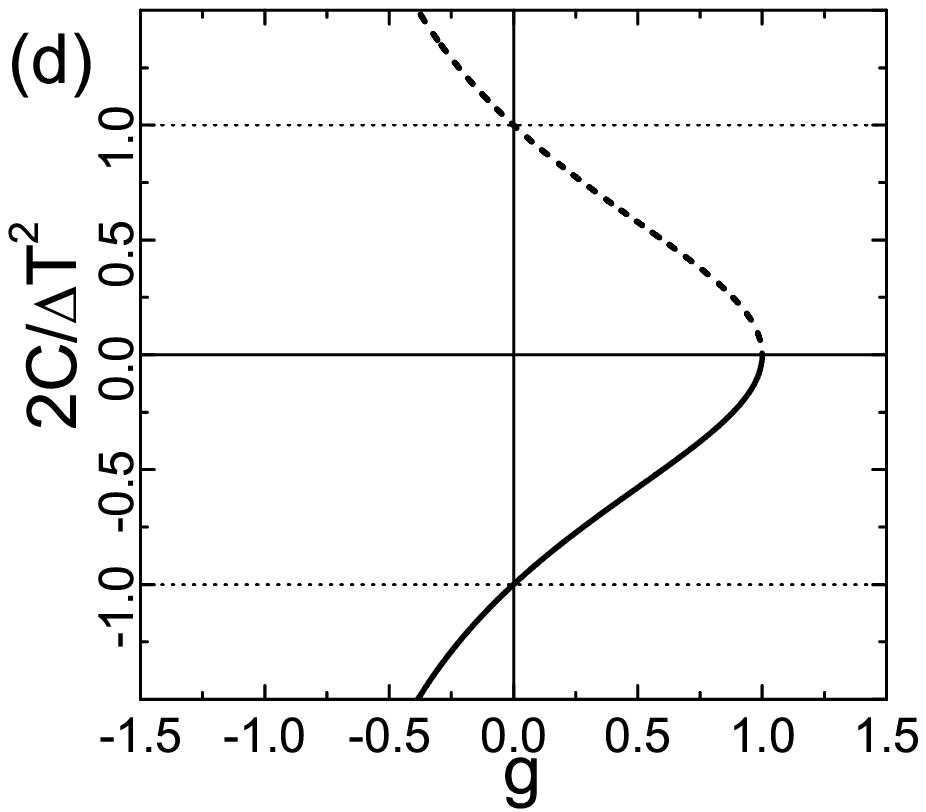}
\caption{\label{Fig3} (a) Chromatic aberration parameter $\gamma$ as a function of the relative chirp $2C/\Delta T^2$ (a) to freeze axially the on-axis CEP at the focal point for different values of $g$ of the input pulse. (b) The same but to have transversally flat CEP at the focal plane. (c,d) Chromatic aberration and chirp to produce on-axis and transversally frozen CEP at focus. Positive chromatic aberration requires negative chirp (solid curves), and vice versa (dashed curves).}
\end{figure}

\begin{figure}[t!]
\includegraphics*[width=8cm]{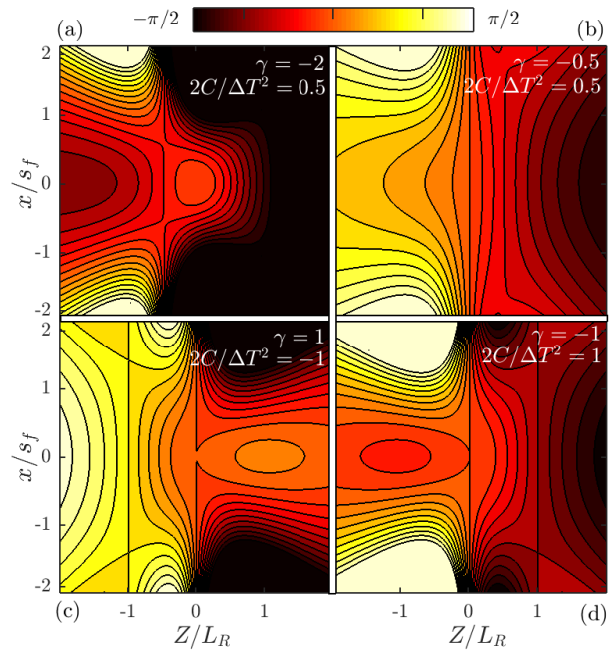}
\caption{\label{Fig4} CEP maps in the focal volume ($y=0$ sections) for $g=0$ and the indicated values of $\gamma$ and $2C/\Delta T^2$. The CEP at $\zeta=0$ is axially constant in (a), transversally constant in (b), axially and transversally constant in (c) and (d), with a maximum in (c) and a minimum in (d).}
\end{figure}

The use of a lens offers more possibilities. According to Eq. (\ref{NOA}), we can have a CEP map with no axial variation at focus with input pulses of any value of $g$ if the chromatic aberration $\gamma$ and the relative chirp $2C/\Delta T^2$ verify
\begin{equation}\label{NOAL}
\gamma \frac{2C}{\Delta T^2} = g - m\,,
\end{equation}
where $m= (1-1/n)^{-1}$. Figure \ref{Fig3}(a) shows the $\gamma-2C/\Delta T^2$ curves determined by Eq. (\ref{NOAL}), limited to relative chirps $|2C/\Delta T^2|<1$ so as not to enlarge excessively the pulse, for different values of its $g$-parameter in the particular case of $n\rightarrow\infty$ ($m=1$), i .e., to freeze the on-axis CEP variation at focus. For other values of $n$ the curves are qualitatively similar. As an example, the particular CEP map for an input pulse with $g=0$ having an on-axis maximum at $\zeta=0$ using the relative chirp $2C/\Delta T^2=0.5$ and the chromatic aberration $\gamma=-2$ is plotted in FIG. \ref{Fig4}(a). With chirp and chromatic aberration of opposite signs, the CEP map would feature a minimum.

Instead, we may wish to have the same CEP at all points of the focal plane. According to Eq. (\ref{NOR}), transversally constant CEP imposes the linear relation
\begin{equation}\label{NORL}
\gamma = -(g+1) \frac{2C}{\Delta T^2}\,.
\end{equation}
The $\gamma-2C/\Delta T^2$ lines determined by Eq. (\ref{NORL}) are depicted in FIG. \ref{Fig3}(b) for the same values of the $g$ parameter as in FIG. \ref{Fig3}(a). The particular CEP map with flat CEP at the focal plane for an input pulse with $g=0$ using relative chirp $2C/\Delta T^2=0.5$ and $\gamma=-0.5$ is plotted in FIG. \ref{Fig4}(b) for illustration purposes. Chirp and chromatic aberration of opposite signs would also produce flat CEP at the focal plane.

While axially and transversally flat CEP at focus with a focusing mirror requires a transform-limited input pulse with a particular value of $g$, with a lens this particularly relevant CEP map is possible with a variety of values of $g$. From Eqs. (\ref{NOAL}) and (\ref{NORL}) we obtain the needed chirp and chromatic aberration as
\begin{equation}\label{FF}
\frac{2C}{\Delta T^2} = \pm \sqrt{\frac{m-g}{1+g}}\,, \quad \gamma =\mp \sqrt{(1+g)(m-g)}\,.
\end{equation}
These values are the points of intersection of the curves in Figures \ref{Fig3}(a) and \ref{Fig3}(b) for each specific value of $g$, intersection points that exist for input pulses with $-1\le g\le m$, and are represented in FIG. \ref{Fig3}(c) and \ref{Fig3}(d) as functions of $g$ in the particular case of $n\rightarrow\infty$ ($m=1$). Positive and negative chromatic aberrations correspond to negative and positive chirp in Eq. (\ref{FF}), or solid and dashed curves in Figures \ref{Fig3}(c) and \ref{Fig3}(d). Thus, given the factor $g$ of the input pulse, there are two specific values of chromatic aberration and chirp to tailor an axially and transversally flat CEP map at focus. Only for $0\le g\le 1$ [black curves in Figures \ref{Fig3}(a) and \ref{Fig3}(b)], however, the chirp can be to be considered small enough according to our criterion of $|2C/\Delta T^2|<1$.
As a couple of examples, the CEP maps corresponding to input pulses characterized by $g=0$ with $\gamma=+1, 2C/\Delta T^2=-1$ (with a local axial maximum at $\zeta=0$) and with $\gamma=-1, 2C/\Delta T^2+1$ (with a local minimum) are plotted in Figures \ref{Fig4}(c) and \ref{Fig4}(d).

\begin{figure*}[t!]
\includegraphics*[width=3.6cm]{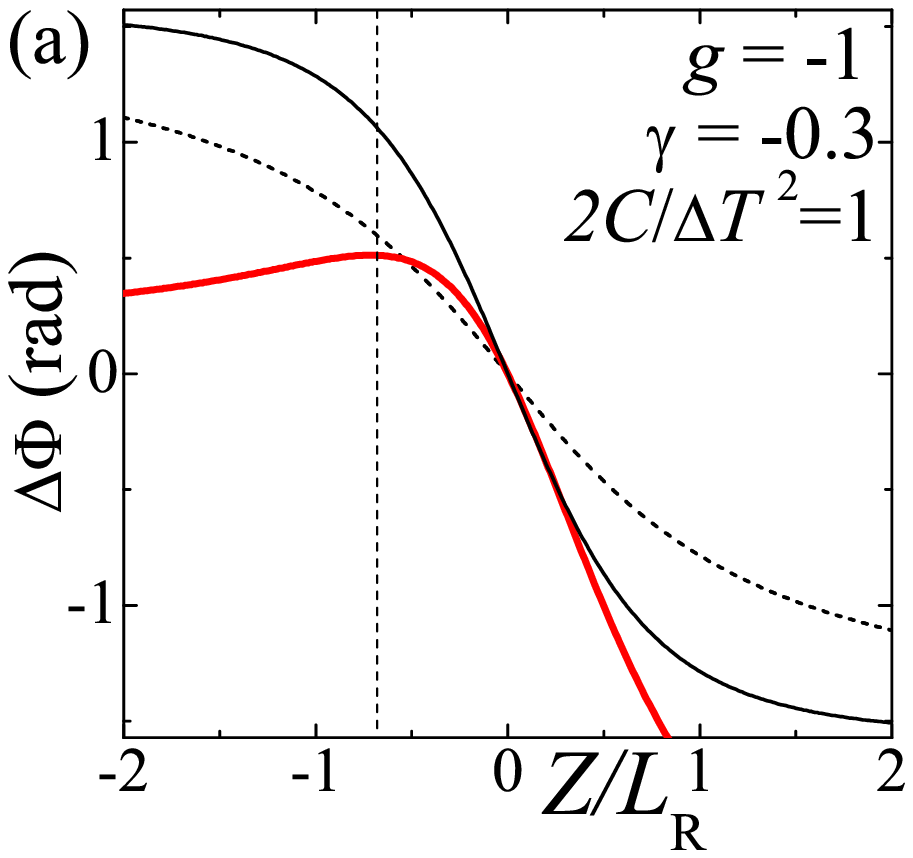}\includegraphics*[width=3.6cm]{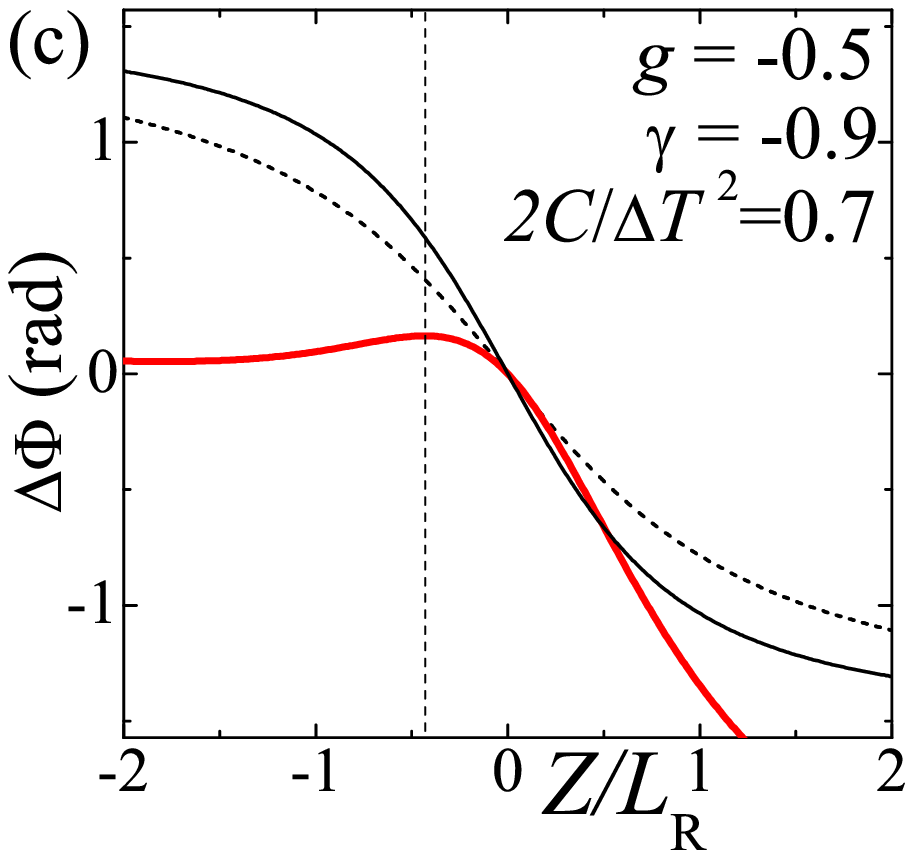}\includegraphics*[width=3.6cm]{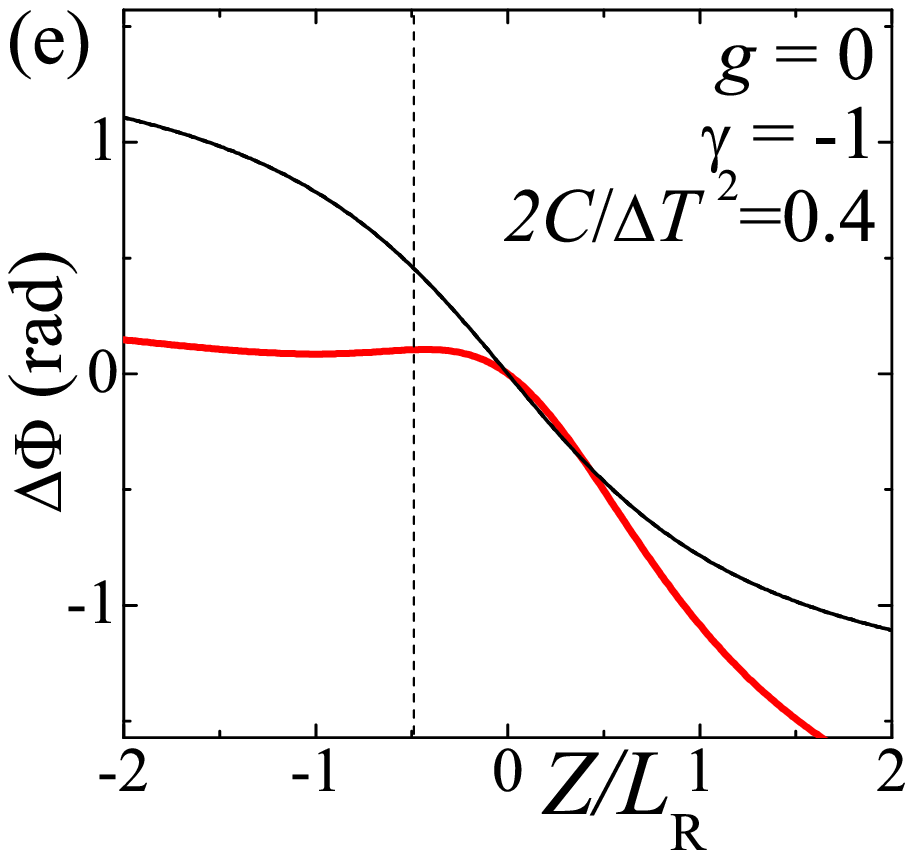}\includegraphics*[width=3.6cm]{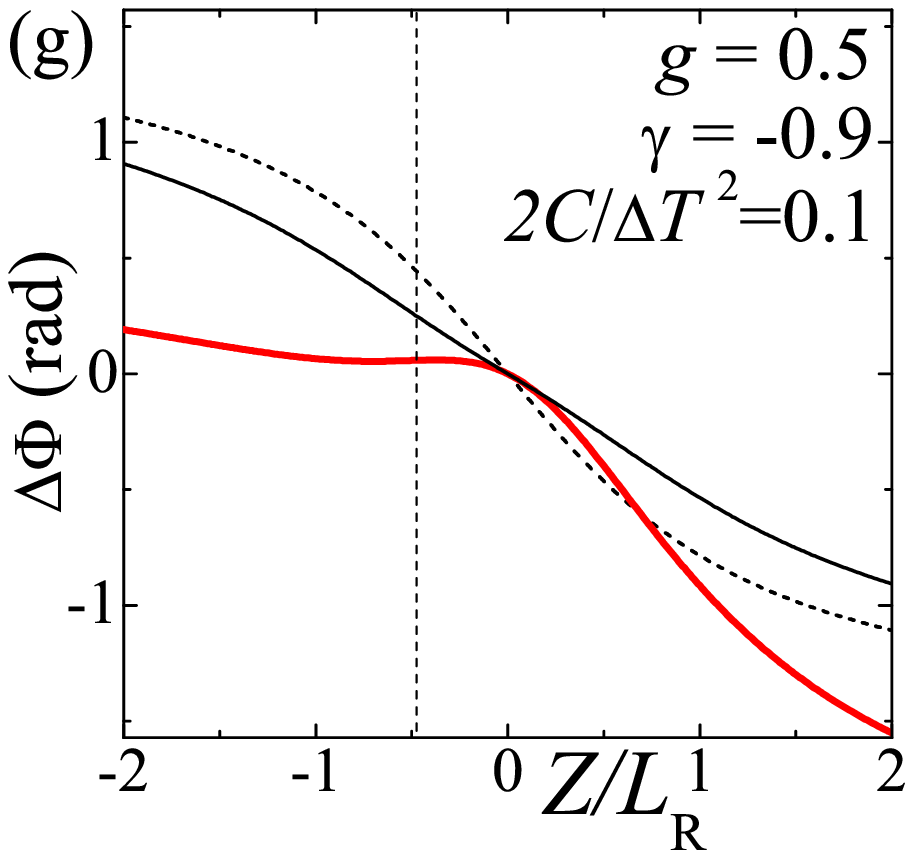}\includegraphics*[width=3.6cm]{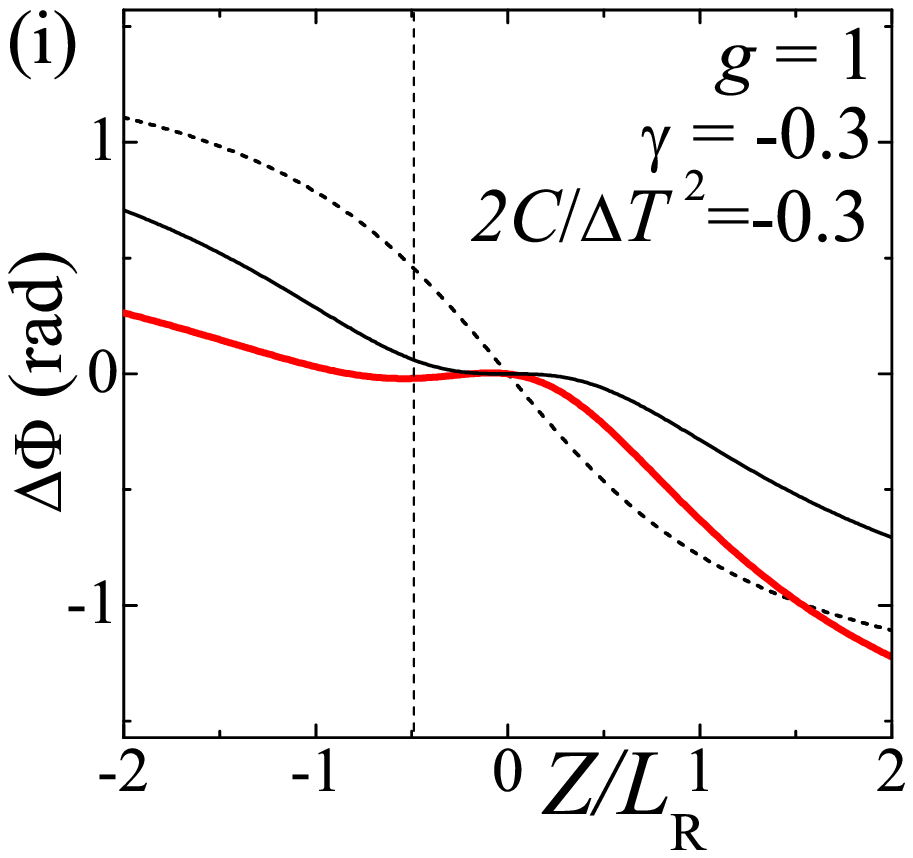}
\includegraphics*[width=3.6cm]{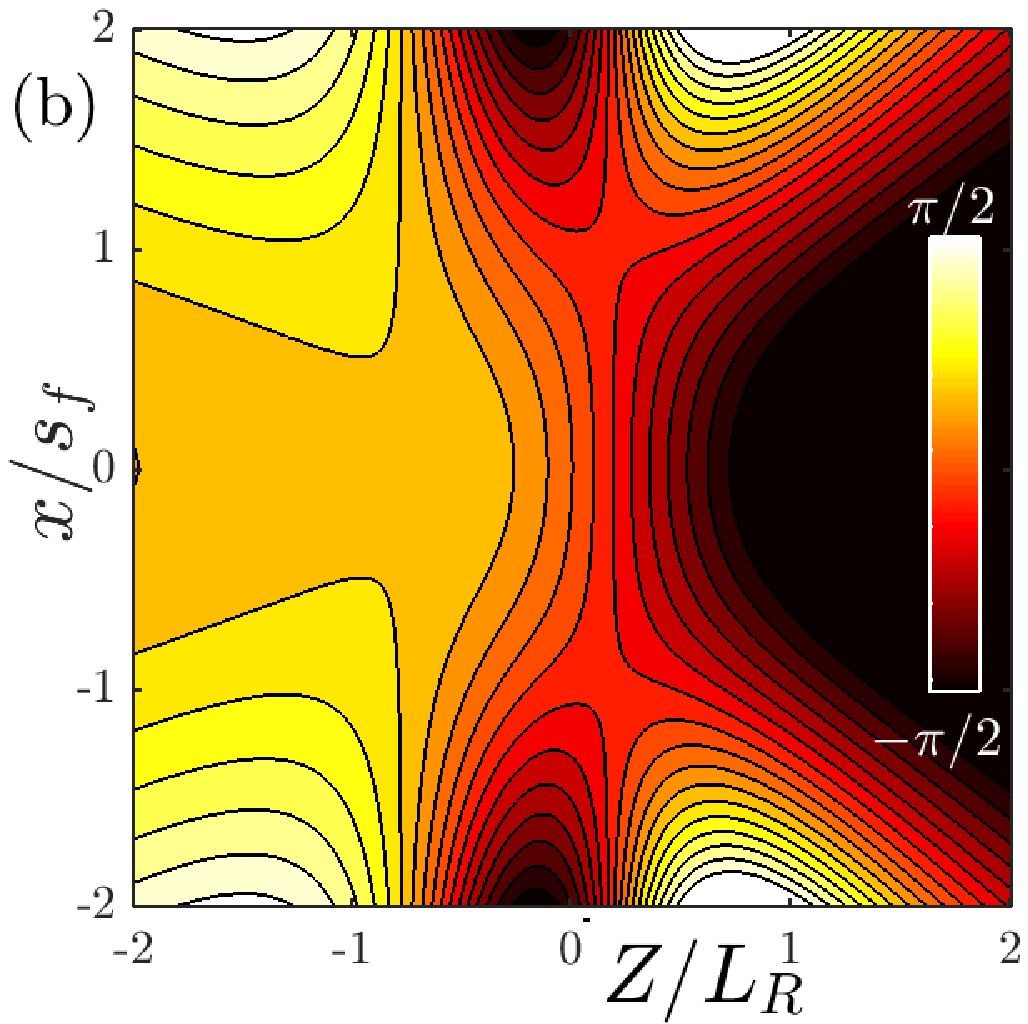}\includegraphics*[width=3.6cm]{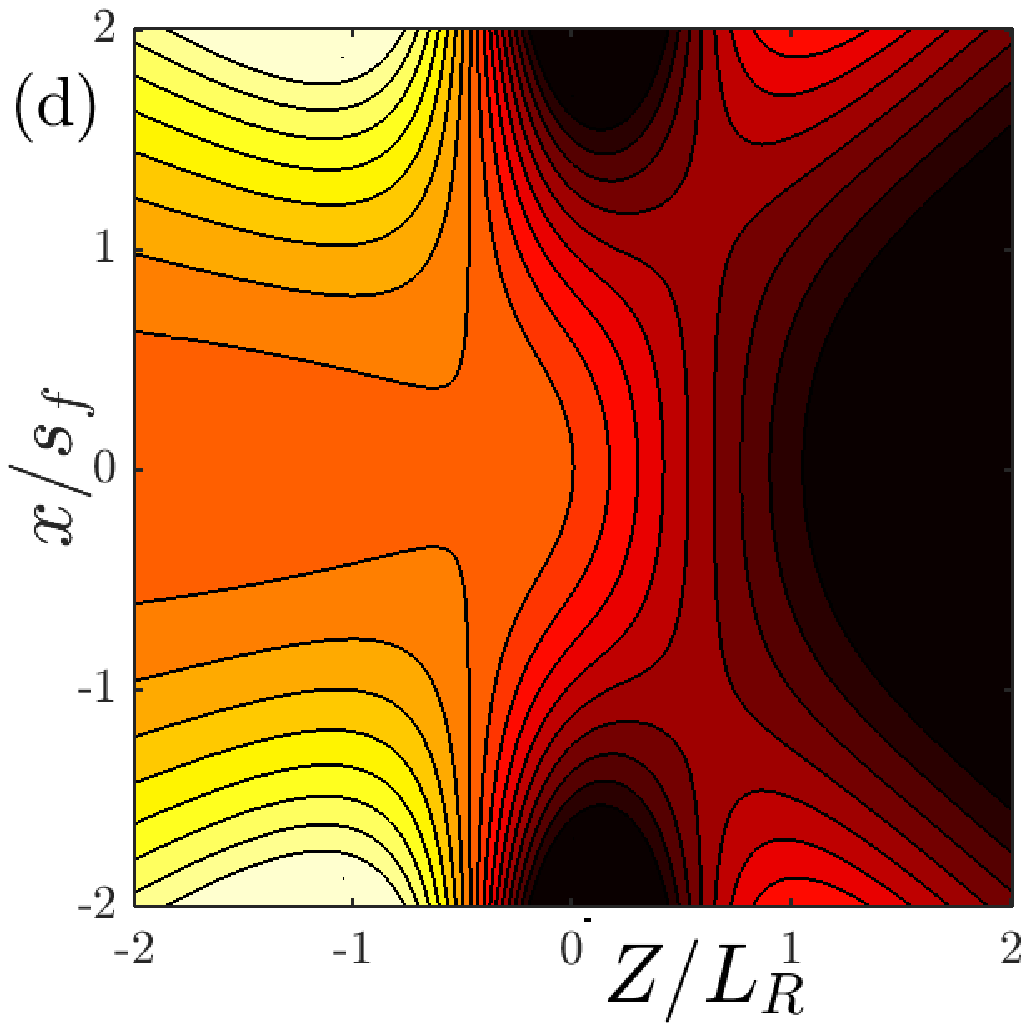}\includegraphics*[width=3.6cm]{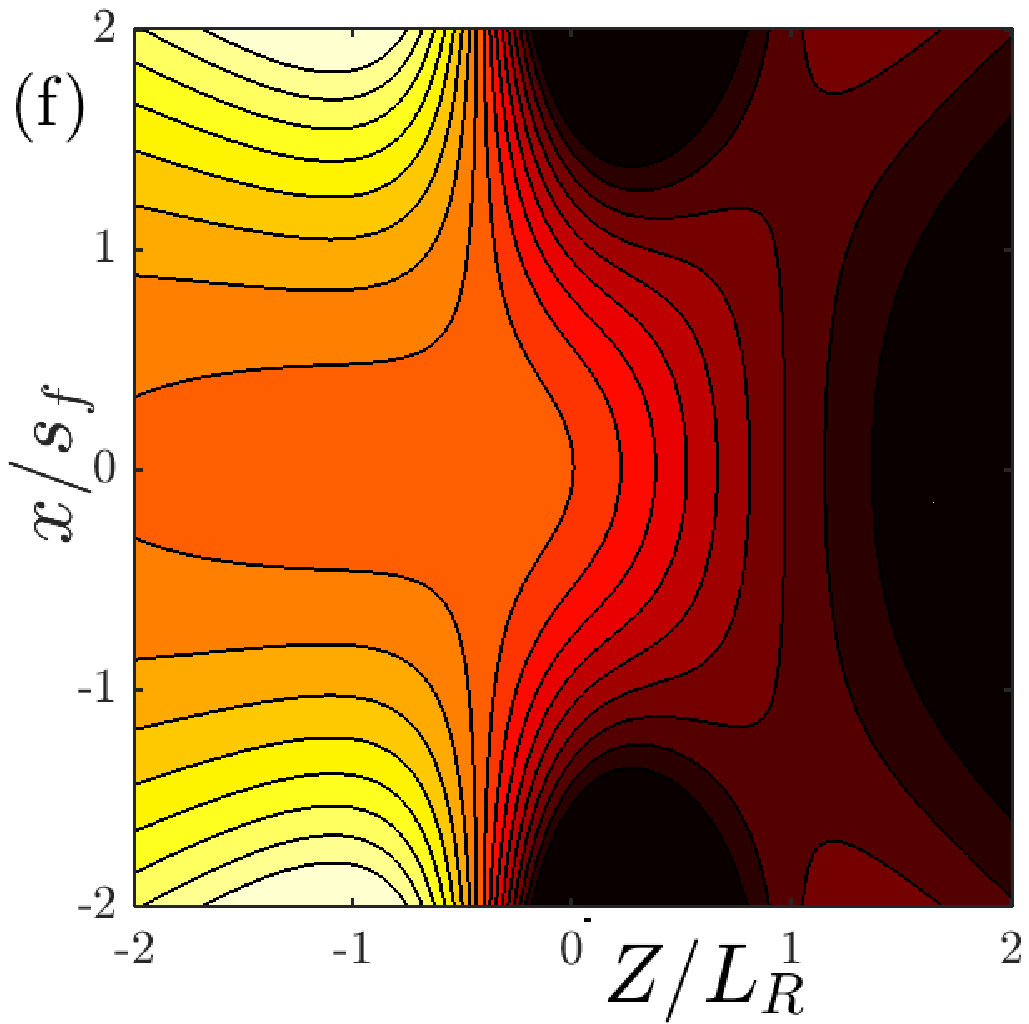}\includegraphics*[width=3.6cm]{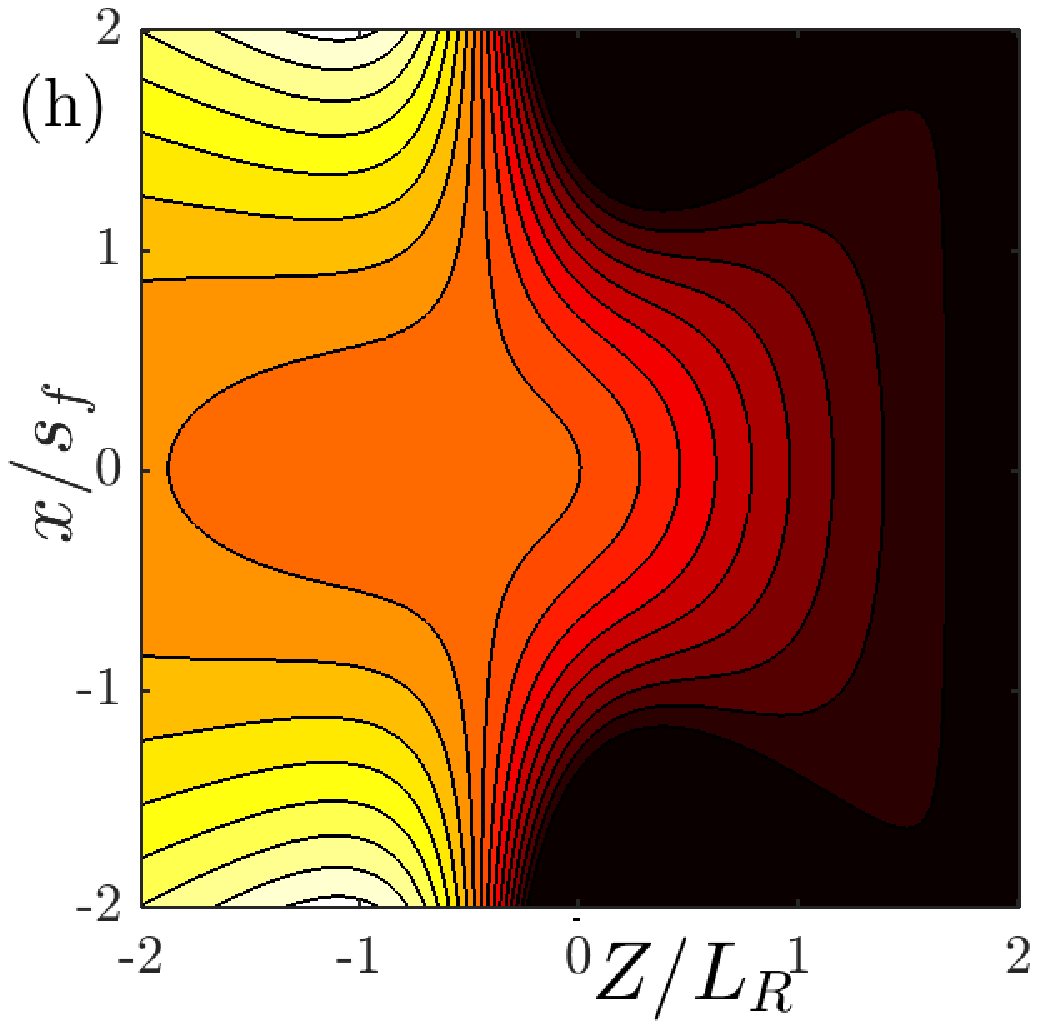}\includegraphics*[width=3.6cm]{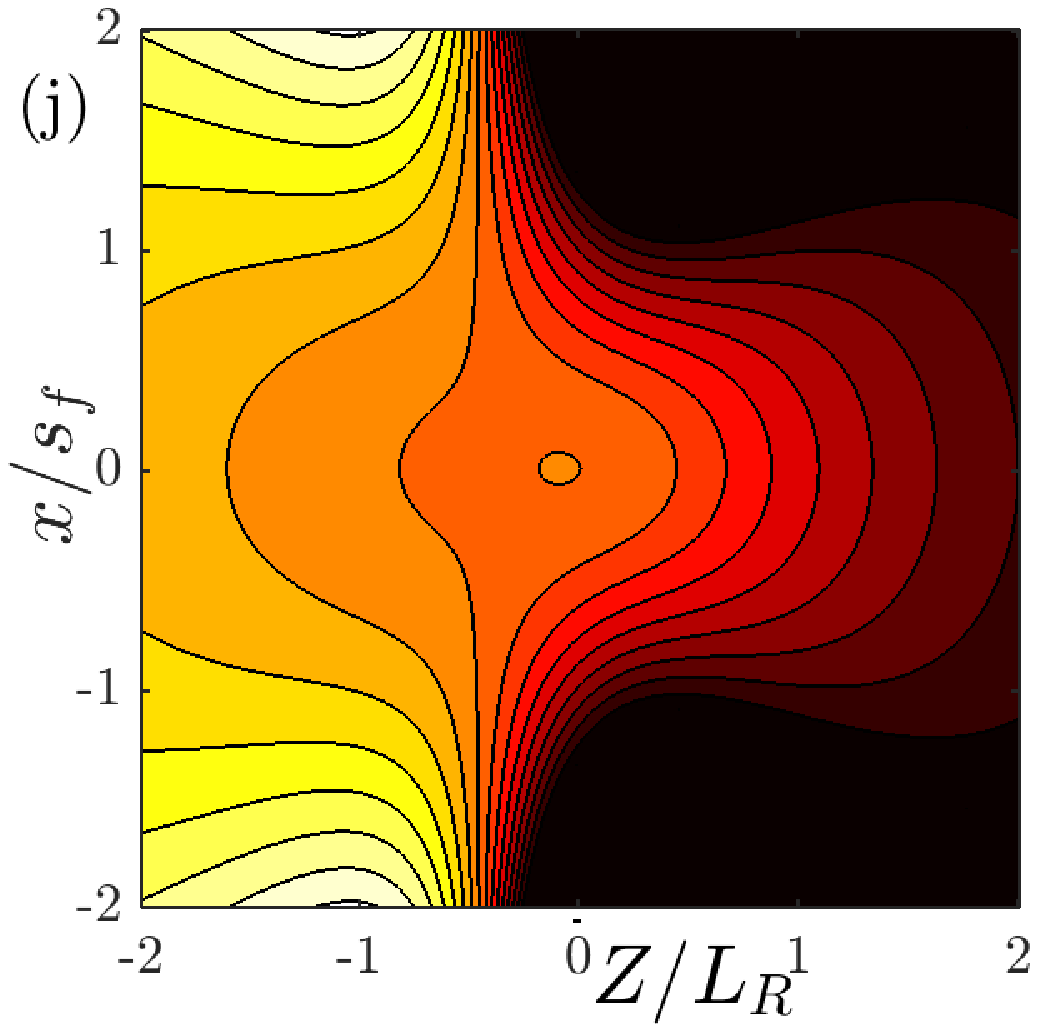}
\caption{\label{Fig5} Upper plots: On-axis ($n\rightarrow\infty$) CEP shift for input pulses with the indicated values of $g$, focused by a lens with the indicated values of chromatic aberration $\gamma$ and chirp $2C/\Delta T^2$ such that the CEP is approximately flat in the first half of the focus (red curves). The dashed curves represent Gouy's phase shift for reference, and the black curves the CEP shift in case of focusing without chromatic aberration and without chirp. Lower plots: Corresponding CEP maps ($y=0$ sections) showing that the CEP is also transversally flat at positions about $-L_R/2$, as indicated by the vertical dashed lines in the upper plots.}
\end{figure*}

Also, it is generally possible to freeze the CEP at particular locations out-of-focus. The required chromatic aberrations can be straightforwardly obtained from Eqs. (\ref{NOR}) and (\ref{NOA}), but the obtained expressions for $\gamma$ and $2C/\Delta T^2$ are long and cumbersome. In Ref. \cite{Porras2012JOSAB}, it has been shown that irrespective of the value of $g$, a small chromatic aberration $\gamma$ in $[-1,0]$  or in $[0,1]$ flattens the axial CEP variation in the first or second half of the focus, respectively, and a small chirp enhances further the flatness of the axial CEP variation. Following this criterion, it is easy in practice, and faster than deriving particular expressions, to adopt a inspection procedure by directly plotting the 3D CEP map given by Eq. (\ref{CEPSHIFTC}) with the given value of $g$, and different values values of $\gamma$ and $2C/\Delta T^2$. We have found that for optimum values of $\gamma$ and $2C/\Delta^2$ to flatten the axial CEP variation in one half of the focus, the transversal CEP variation at about the middle of that half vanishes. This result is illustrated in FIG. \ref{Fig5} for focused input pulses with different values of $g$, including negative ones. The vertical dashed lines in these figures indicate the positions about the middle of the first half of the focus where the CEP is axially and transversally flat. For each $g$, chromatic aberrations and chirp of opposite signs would equally flatten axially and transversally the CEP in the second half of the focus.

\section{Implementation of tunable chromatic aberration for CEP map control}\label{sec:doublet}

In addition to small chirps $|2C/\Delta T^2|\le 1$, we propose a simple "separable" close-to-achromatic doublet to easily tune the chromatic aberration parameter $\gamma$, and hence to focus few-cycle pulses with tailored CEP map. Substantial chromatic aberration can severely distort the pulse \cite{Bor1992OC}, but condition $|\gamma|< \omega_0/\Delta\omega$ ensures negligible pulse distortion, and according to the above analysis $|\gamma|\le 1$ generally suffices to the purpose of CEP map design.

For two thin lenses of focal lengths $f_1$ and $f_2$  separated a distance $d$, the focal length is
\begin{equation}\label{eq:ftot}
\frac{1}{f} = \frac{1}{f_{1}} + \frac{1}{f_{2}} - \frac{d}{f_{1} f_{2}}\,.
\end{equation}
Lensmaker's formula for thin lenses and Eq. (\ref{eq:g_gamma}) yields
\begin{equation} \label{eq:gamma_lenssystem}
\gamma = -L \left[ \frac{a}{f_{1}} + \frac{b}{f_{2}} - \frac{a+b}{f_{1} f_{2}} d \right] = \gamma(0) + L \frac{a+b}{f_1f_2}d\,,
\end{equation}
where $a = n'_{1} \omega/(n_{1} -1)$ and $b = n'_{2} \omega/(n_{2} -1)$, $n_{1,2}$ are the lens refractive indexes, $L=\omega s^2/2c=f^2/L_R$ is the Rayleigh distance of the input pulse, and, as above, all quantities and their derivatives are evaluated at $\omega_0$. In the second equality $\gamma(0)=-L(a/f_1 + b/f_2)$ is the chromatic aberration parameter when the two lenses are joint. Thus, chromatic aberration varies linearly with increasing separation $d$, giving an easy way to control it and hence the CEP map. Since usually achromatic doublets consist of a focusing and a defocusing element, meaning $f_{1} f_{2} < 0$, $\gamma$ decreases with lens separation.

Suppose we wish to focus the input pulse of the Rayleigh distance $L$ to a half-focal depth $L_R$, as required by a particular experimental setup, and with chromatic aberration $\gamma(0)$ when the two lenses are joint. Setting the required focal length $\sqrt{LL_R}$ in Eq. (\ref{eq:ftot}) with $d=0$ and from $\gamma(0)=-L(a/f_1 + b/f_2)$, we obtain the required focal lengths as
\begin{equation}\label{eq:ad_gorig}
\frac{1}{f_{1}} = \frac{\gamma(0)/L + b/\sqrt{L L_{R}}}{b-a}\,, \quad \frac{1}{f_{2}} = \frac{-\gamma(0)/L - a/\sqrt{L L_{R}}}{b-a}\,.
\end{equation}
By slightly separating the two lenses, the chromatic aberration $\gamma$ can be tuned following the second equality in Eq. (\ref{eq:gamma_lenssystem}), and if the system is designed such that the required separations $d$ for efficient $\gamma$ tuning are minimal compared to $f_1$ and $f_2$, the variation of the total focal length $f$ is also minimal, and the variation of the focused Rayleigh distance $L_R= f^2/L$ is also very small.

As an example of design, we consider pulses at 800 nm carrier wavelength, input spot size $s_L=17$ mm (20 mm FWHM in intensity, and $L=1.135\times 10^6$ mm) to be focused to $L_R=3$ mm. Chromatic aberration is wished to be tuned from $\gamma=+1$ to $\gamma=-1$. With a N-BAF10 lens and a SF10 lens, Eqs. (\ref{eq:ad_gorig}) with $\gamma(0)=+1$ give $f_{1} = 589.2\,\mathrm{mm}$ and $f_{2} = -867.9\,\mathrm{mm}$, and the total focal length with zero separation is $f=184.5$ cm. From Eq. (\ref{eq:gamma_lenssystem}), $\gamma$ varies linearly ranges from $+1$ to $-1$ when $d$ ranges from $d=0$ to $d=10$ mm. At the same time, the total focal length obtained from Eq. (\ref{eq:ftot}) experiences a small variation from $184.5\,\mathrm{cm}$ to $178.1\,\mathrm{cm}$, corresponding to a negligible decrease of the focused Rayleigh range $L_{R}$ from $3\,\mathrm{mm}$ to $2.8\,\mathrm{mm}$. Due to the small change of the focal length the doublet can be placed and moved on a linear stage, so the target remains at the same position in the focal volume. This means that the proposed doublet system, along with small chirps of the focused pulse, allows one to study light-matter interactions with practically the same parameters except the CEP map, that can be tuned to meet the needs of the specific experiment.

\begin{figure}[tb!]
\includegraphics*[width=4.3cm]{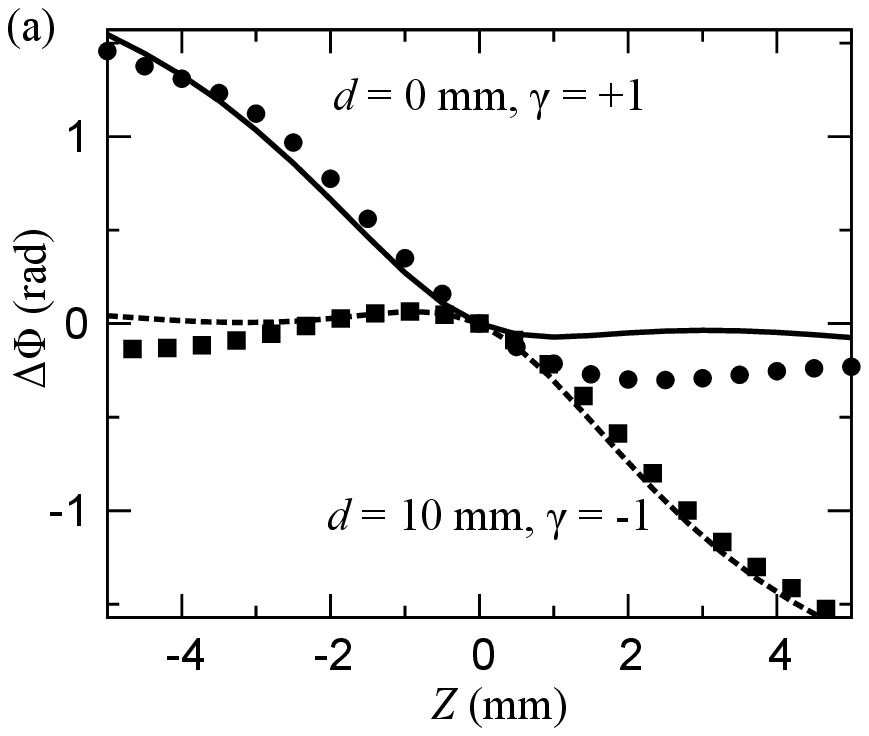}\includegraphics*[width=4.3cm]{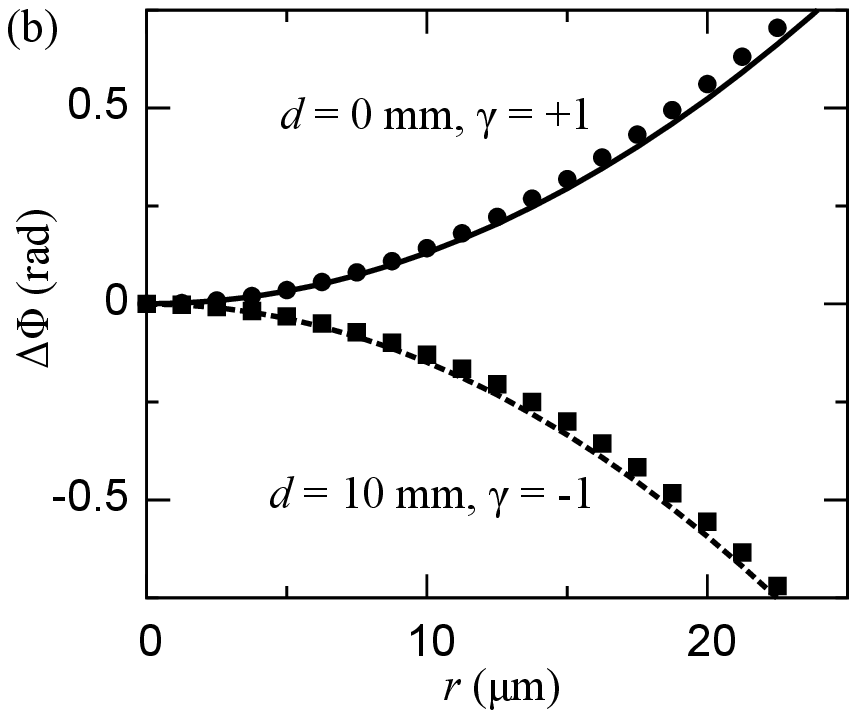}
\caption{\label{Fig6} (a) On-axis CEP shift about the focus for two different lens separation distances evaluated from Eq. (\ref{CEPSHIFTC}) (curves) and from the numerical simulation described in the text (symbols). (b) The same but at the focal plane.}
\end{figure}

We have confirmed numerically the usefulness of the above system for focusing without distortion few-cycle pulses. Similarly to previous works \cite{Porras2012APB, Porras2012JOSAB, MajorPhD}, our numerical calculations use a ray-tracing algorithm to analyze propagation from the input to the output plane of the focusing optics, while the electric field strengths in specific points in the focal volume are evaluated from scalar diffraction theory. The model takes into account the truncation of the beam by the aperture of the optics, and considers the lens varying thickness along with the resulting spherical and chromatic aberrations \cite{MajorPhD}. The above system is realized with central lens thicknesses $D_1=3.5$ mm, $D_2=2.0$ mm, $3$ inch diameter, radii of curvature $120\,\mathrm{cm}$ and $58\,\mathrm{cm}$ for the first lens, and $-58\,\mathrm{cm}$ and $881\,\mathrm{cm}$ for the second lens, making it possible zero separation (negative value means a concave surface). In the ray-tracing calculations, $d$ is assimilated to the distance between the back and front surfaces of the two lenses. The input pulsed beam at $800$ nm carrier wavelength is Gaussian temporally and transversally, of duration $\Delta T= 8.49$ fs ($10$ fs FWHM), width $s_L=17$ mm ($20$ mm FWHM), and $g$ factor equal to $0.5$. With precompensation of the second and third order dispersion introduced by the central thicknesses $D_1$ and $D_2$, the pulse after the lenses is Gaussian and transform-limited to all practical purposes and of the original duration at all points of the focal volume, similarly to the examples of previous works \cite{Porras2012APB, Porras2012JOSAB, Major2015AO, MajorPhD} where the usefulness of lenses to focus few-cycle pulses is stressed. The symbols in Figures \ref{Fig6}(a) and \ref{Fig6}(b) represent the CEP shift along the beam axis and at along the focal plane, respectively, extracted from the numerically evaluated electric fields when the lens separation is $d=0$ mm and $d=10$, and the solid curves represent the predictions of Eq. (\ref{CEPSHIFT}) with $g=0.5$, $C=0$ and $\gamma=\pm 1$, demonstrating also the usefulness of the separable doublet system for CEP map tailoring. Additional calculations confirm applicability of the system with pulse durations down to $5\,\mathrm{fs}$ (two-cycle pulse). With shorter pulses, precompensation of higher order dispersion terms than the second and third might be necessary to retain the pulse shape.

\section{Conclusions}

We have completed partial calculations provided by different authors in previous works in order to provide the complete spatial distribution of the CEP of few-cycle, Gaussian-Gaussian pulsed beams focused by mirrors or lenses. This amounts to specify the complete electric field under the envelope at each position of the focal volume. Our results underscore the importance of characterizing the few-cycle laser source in use by measuring its factor $g$, since it is the fundamental parameter that specifies the CEP map in the focal volume. Once $g$ is determined, small residual chirps and small chromatic aberration can be used to modify and adapt the CEP map for specific applications or target positions without appreciably deteriorating the pulse. We have shown, for example, how to suppress the CEP variation both longitudinally and transversally at different locations of the focal volume. We have also proposed and tested numerically a quasi-achromatic doublet that can perform this job without distorting the few-cycle pulse, and only changing the CEP map. This analysis, or others more adapted to each particular situation, can help to reach conditions that are beneficial in research areas where light-matter interactions strongly depend on the CEP, such as high-harmonic generation, attosecond pulse production, ultrafast nanooptics or plasmonics. As possible extensions of this work, it would be of interest to remove the restriction to Gaussian beams in order to evaluate the CEP map of other focused beams in use in these phase-sensitive light-matter interactions, e. g., of vortex beams \cite{Gauthier2017Nature} or radially/azimuthally polarized beams \cite{Biss2003OL}.

\section*{Acknowledgments}

Projects of the Spanish Ministerio de Econom\'{\i}a y Competitividad No. MTM2015-63914-P and No. FIS2017-87360-P. The ELI-ALPS project (GINOP-2.3.6-15-2015-00001) is supported by the European Union and co-financed by the European Regional Development Fund.

\bibliography{references_CEPmap}

\begin{thebibliography}{35}%
\makeatletter
\providecommand \@ifxundefined [1]{%
 \@ifx{#1\undefined}
}%
\providecommand \@ifnum [1]{%
 \ifnum #1\expandafter \@firstoftwo
 \else \expandafter \@secondoftwo
 \fi
}%
\providecommand \@ifx [1]{%
 \ifx #1\expandafter \@firstoftwo
 \else \expandafter \@secondoftwo
 \fi
}%
\providecommand \natexlab [1]{#1}%
\providecommand \enquote  [1]{``#1''}%
\providecommand \bibnamefont  [1]{#1}%
\providecommand \bibfnamefont [1]{#1}%
\providecommand \citenamefont [1]{#1}%
\providecommand \href@noop [0]{\@secondoftwo}%
\providecommand \href [0]{\begingroup \@sanitize@url \@href}%
\providecommand \@href[1]{\@@startlink{#1}\@@href}%
\providecommand \@@href[1]{\endgroup#1\@@endlink}%
\providecommand \@sanitize@url [0]{\catcode `\\12\catcode `\$12\catcode
  `\&12\catcode `\#12\catcode `\^12\catcode `\_12\catcode `\%12\relax}%
\providecommand \@@startlink[1]{}%
\providecommand \@@endlink[0]{}%
\providecommand \url  [0]{\begingroup\@sanitize@url \@url }%
\providecommand \@url [1]{\endgroup\@href {#1}{\urlprefix }}%
\providecommand \urlprefix  [0]{URL }%
\providecommand \Eprint [0]{\href }%
\providecommand \doibase [0]{http://dx.doi.org/}%
\providecommand \selectlanguage [0]{\@gobble}%
\providecommand \bibinfo  [0]{\@secondoftwo}%
\providecommand \bibfield  [0]{\@secondoftwo}%
\providecommand \translation [1]{[#1]}%
\providecommand \BibitemOpen [0]{}%
\providecommand \bibitemStop [0]{}%
\providecommand \bibitemNoStop [0]{.\EOS\space}%
\providecommand \EOS [0]{\spacefactor3000\relax}%
\providecommand \BibitemShut  [1]{\csname bibitem#1\endcsname}%
\let\auto@bib@innerbib\@empty
\bibitem [{\citenamefont {Krausz}\ and\ \citenamefont
  {Stockman}(2014)}]{Krausz2014NPhot}%
  \BibitemOpen
  \bibfield  {author} {\bibinfo {author} {\bibfnamefont {F.}~\bibnamefont
  {Krausz}}\ and\ \bibinfo {author} {\bibfnamefont {M.~I.}\ \bibnamefont
  {Stockman}},\ }\href {\doibase 10.1038/nphoton.2014.28} {\bibfield  {journal}
  {\bibinfo  {journal} {Nat. Photonics}\ }\textbf {\bibinfo {volume} {8}},\
  \bibinfo {pages} {205} (\bibinfo {year} {2014})}\BibitemShut {NoStop}%
\bibitem [{\citenamefont {Baltuska}\ \emph {et~al.}(2003)\citenamefont
  {Baltuska}, \citenamefont {Udem}, \citenamefont {Uiberacker}, \citenamefont
  {Hentschel}, \citenamefont {Goulielmakis}, \citenamefont {Gohle},
  \citenamefont {Holzwarth}, \citenamefont {Yakovlev}, \citenamefont {Scrinzi},
  \citenamefont {Hansch},\ and\ \citenamefont {Krausz}}]{Baltuska2003Nature}%
  \BibitemOpen
  \bibfield  {author} {\bibinfo {author} {\bibfnamefont {A.}~\bibnamefont
  {Baltuska}}, \bibinfo {author} {\bibfnamefont {T.}~\bibnamefont {Udem}},
  \bibinfo {author} {\bibfnamefont {M.}~\bibnamefont {Uiberacker}}, \bibinfo
  {author} {\bibfnamefont {M.}~\bibnamefont {Hentschel}}, \bibinfo {author}
  {\bibfnamefont {E.}~\bibnamefont {Goulielmakis}}, \bibinfo {author}
  {\bibfnamefont {C.}~\bibnamefont {Gohle}}, \bibinfo {author} {\bibfnamefont
  {R.}~\bibnamefont {Holzwarth}}, \bibinfo {author} {\bibfnamefont
  {V.}~\bibnamefont {Yakovlev}}, \bibinfo {author} {\bibfnamefont
  {A.}~\bibnamefont {Scrinzi}}, \bibinfo {author} {\bibfnamefont
  {T.}~\bibnamefont {Hansch}}, \ and\ \bibinfo {author} {\bibfnamefont
  {F.}~\bibnamefont {Krausz}},\ }\href {\doibase 10.1038/nature01483}
  {\bibfield  {journal} {\bibinfo  {journal} {Nature}\ }\textbf {\bibinfo
  {volume} {422}},\ \bibinfo {pages} {189} (\bibinfo {year}
  {2003})}\BibitemShut {NoStop}%
\bibitem [{\citenamefont {Haworth}\ \emph {et~al.}(2007)\citenamefont
  {Haworth}, \citenamefont {Chipperfield}, \citenamefont {Robinson},
  \citenamefont {Knight}, \citenamefont {Marangos},\ and\ \citenamefont
  {Tisch}}]{Haworth2007NPhys}%
  \BibitemOpen
  \bibfield  {author} {\bibinfo {author} {\bibfnamefont {C.~A.}\ \bibnamefont
  {Haworth}}, \bibinfo {author} {\bibfnamefont {L.~E.}\ \bibnamefont
  {Chipperfield}}, \bibinfo {author} {\bibfnamefont {J.~S.}\ \bibnamefont
  {Robinson}}, \bibinfo {author} {\bibfnamefont {P.~L.}\ \bibnamefont
  {Knight}}, \bibinfo {author} {\bibfnamefont {J.~P.}\ \bibnamefont
  {Marangos}}, \ and\ \bibinfo {author} {\bibfnamefont {J.~W.~G.}\ \bibnamefont
  {Tisch}},\ }\href {\doibase 10.1038/nphys463} {\bibfield  {journal} {\bibinfo
   {journal} {Nat. Phys.}\ }\textbf {\bibinfo {volume} {3}},\ \bibinfo {pages}
  {52} (\bibinfo {year} {2007})}\BibitemShut {NoStop}%
\bibitem [{\citenamefont {Ishii}\ \emph {et~al.}(2014)\citenamefont {Ishii},
  \citenamefont {Kaneshima}, \citenamefont {Kitano}, \citenamefont {Kanai},
  \citenamefont {Watanabe},\ and\ \citenamefont {Itatani}}]{Ishii2014NC}%
  \BibitemOpen
  \bibfield  {author} {\bibinfo {author} {\bibfnamefont {N.}~\bibnamefont
  {Ishii}}, \bibinfo {author} {\bibfnamefont {K.}~\bibnamefont {Kaneshima}},
  \bibinfo {author} {\bibfnamefont {K.}~\bibnamefont {Kitano}}, \bibinfo
  {author} {\bibfnamefont {T.}~\bibnamefont {Kanai}}, \bibinfo {author}
  {\bibfnamefont {S.}~\bibnamefont {Watanabe}}, \ and\ \bibinfo {author}
  {\bibfnamefont {J.}~\bibnamefont {Itatani}},\ }\href {\doibase
  10.1038/ncomms4331} {\bibfield  {journal} {\bibinfo  {journal} {Nat.
  Commun.}\ }\textbf {\bibinfo {volume} {5}},\ \bibinfo {pages} {3331}
  (\bibinfo {year} {2014})}\BibitemShut {NoStop}%
\bibitem [{\citenamefont {Kr\"uger}\ \emph {et~al.}(2011)\citenamefont
  {Kr\"uger}, \citenamefont {Schenk},\ and\ \citenamefont
  {Hommelhoff}}]{Kruger2011Nature}%
  \BibitemOpen
  \bibfield  {author} {\bibinfo {author} {\bibfnamefont {M.}~\bibnamefont
  {Kr\"uger}}, \bibinfo {author} {\bibfnamefont {M.}~\bibnamefont {Schenk}}, \
  and\ \bibinfo {author} {\bibfnamefont {P.}~\bibnamefont {Hommelhoff}},\
  }\href {\doibase 10.1038/nature10196} {\bibfield  {journal} {\bibinfo
  {journal} {Nature}\ }\textbf {\bibinfo {volume} {475}},\ \bibinfo {pages}
  {78} (\bibinfo {year} {2011})}\BibitemShut {NoStop}%
\bibitem [{\citenamefont {Chini}\ \emph {et~al.}(2014)\citenamefont {Chini},
  \citenamefont {Zhao},\ and\ \citenamefont {Chang}}]{Chini2014NPhot}%
  \BibitemOpen
  \bibfield  {author} {\bibinfo {author} {\bibfnamefont {M.}~\bibnamefont
  {Chini}}, \bibinfo {author} {\bibfnamefont {K.}~\bibnamefont {Zhao}}, \ and\
  \bibinfo {author} {\bibfnamefont {Z.}~\bibnamefont {Chang}},\ }\href
  {\doibase 10.1038/nphoton.2013.362} {\bibfield  {journal} {\bibinfo
  {journal} {Nat. Photonics}\ }\textbf {\bibinfo {volume} {8}},\ \bibinfo
  {pages} {178} (\bibinfo {year} {2014})}\BibitemShut {NoStop}%
\bibitem [{\citenamefont {Popmintchev}\ \emph {et~al.}(2010)\citenamefont
  {Popmintchev}, \citenamefont {Chen}, \citenamefont {Arpin}, \citenamefont
  {Murnane},\ and\ \citenamefont {Kapteyn}}]{Popmintchev2010NPhot}%
  \BibitemOpen
  \bibfield  {author} {\bibinfo {author} {\bibfnamefont {T.}~\bibnamefont
  {Popmintchev}}, \bibinfo {author} {\bibfnamefont {M.-C.}\ \bibnamefont
  {Chen}}, \bibinfo {author} {\bibfnamefont {P.}~\bibnamefont {Arpin}},
  \bibinfo {author} {\bibfnamefont {M.~M.}\ \bibnamefont {Murnane}}, \ and\
  \bibinfo {author} {\bibfnamefont {H.~C.}\ \bibnamefont {Kapteyn}},\ }\href
  {\doibase 10.1038/nphoton.2010.256} {\bibfield  {journal} {\bibinfo
  {journal} {Nature Photon.}\ }\textbf {\bibinfo {volume} {4}},\ \bibinfo
  {pages} {822} (\bibinfo {year} {2010})}\BibitemShut {NoStop}%
\bibitem [{\citenamefont {Rudawski}\ \emph {et~al.}(2015)\citenamefont
  {Rudawski}, \citenamefont {Harth}, \citenamefont {Guo}, \citenamefont
  {Lorek}, \citenamefont {Miranda}, \citenamefont {Heyl}, \citenamefont
  {Larsen}, \citenamefont {Ahrens}, \citenamefont {Prochnow}, \citenamefont
  {Binhammer}, \citenamefont {Morgner}, \citenamefont {Mauritsson},
  \citenamefont {L’Huillier},\ and\ \citenamefont
  {Arnold}}]{Rudawski2015EPJD}%
  \BibitemOpen
  \bibfield  {author} {\bibinfo {author} {\bibfnamefont {P.}~\bibnamefont
  {Rudawski}}, \bibinfo {author} {\bibfnamefont {A.}~\bibnamefont {Harth}},
  \bibinfo {author} {\bibfnamefont {C.}~\bibnamefont {Guo}}, \bibinfo {author}
  {\bibfnamefont {E.}~\bibnamefont {Lorek}}, \bibinfo {author} {\bibfnamefont
  {M.}~\bibnamefont {Miranda}}, \bibinfo {author} {\bibfnamefont {C.~M.}\
  \bibnamefont {Heyl}}, \bibinfo {author} {\bibfnamefont {E.~W.}\ \bibnamefont
  {Larsen}}, \bibinfo {author} {\bibfnamefont {J.}~\bibnamefont {Ahrens}},
  \bibinfo {author} {\bibfnamefont {O.}~\bibnamefont {Prochnow}}, \bibinfo
  {author} {\bibfnamefont {T.}~\bibnamefont {Binhammer}}, \bibinfo {author}
  {\bibfnamefont {U.}~\bibnamefont {Morgner}}, \bibinfo {author} {\bibfnamefont
  {J.}~\bibnamefont {Mauritsson}}, \bibinfo {author} {\bibfnamefont
  {A.}~\bibnamefont {L’Huillier}}, \ and\ \bibinfo {author} {\bibfnamefont
  {C.~L.}\ \bibnamefont {Arnold}},\ }\href {\doibase
  10.1140/epjd/e2015-50568-y} {\bibfield  {journal} {\bibinfo  {journal} {Eur.
  Phys. J. D}\ }\textbf {\bibinfo {volume} {69}},\ \bibinfo {eid} {70}
  (\bibinfo {year} {2015})}\BibitemShut {NoStop}%
\bibitem [{\citenamefont {Tritschler}\ \emph {et~al.}(2005)\citenamefont
  {Tritschler}, \citenamefont {Hof}, \citenamefont {Klein},\ and\ \citenamefont
  {Wegener}}]{Tritschler2005OL}%
  \BibitemOpen
  \bibfield  {author} {\bibinfo {author} {\bibfnamefont {T.}~\bibnamefont
  {Tritschler}}, \bibinfo {author} {\bibfnamefont {K.~D.}\ \bibnamefont {Hof}},
  \bibinfo {author} {\bibfnamefont {M.~W.}\ \bibnamefont {Klein}}, \ and\
  \bibinfo {author} {\bibfnamefont {M.}~\bibnamefont {Wegener}},\ }\href
  {\doibase 10.1364/OL.30.000753} {\bibfield  {journal} {\bibinfo  {journal}
  {Opt. Lett.}\ }\textbf {\bibinfo {volume} {30}},\ \bibinfo {pages} {753}
  (\bibinfo {year} {2005})}\BibitemShut {NoStop}%
\bibitem [{\citenamefont {Yang}\ \emph {et~al.}(2008)\citenamefont {Yang},
  \citenamefont {Yang},\ and\ \citenamefont {Zhang}}]{Yang2008COL}%
  \BibitemOpen
  \bibfield  {author} {\bibinfo {author} {\bibfnamefont {Z.}~\bibnamefont
  {Yang}}, \bibinfo {author} {\bibfnamefont {Z.}~\bibnamefont {Yang}}, \ and\
  \bibinfo {author} {\bibfnamefont {S.}~\bibnamefont {Zhang}},\ }\href
  {http://col.osa.org/abstract.cfm?URI=col-6-3-189} {\bibfield  {journal}
  {\bibinfo  {journal} {Chin. Opt. Lett.}\ }\textbf {\bibinfo {volume} {6}},\
  \bibinfo {pages} {189} (\bibinfo {year} {2008})}\BibitemShut {NoStop}%
\bibitem [{\citenamefont {Major}\ \emph
  {et~al.}(2015{\natexlab{a}})\citenamefont {Major}, \citenamefont {Horváth},\
  and\ \citenamefont {Porras}}]{Major2015JO}%
  \BibitemOpen
  \bibfield  {author} {\bibinfo {author} {\bibfnamefont {B.}~\bibnamefont
  {Major}}, \bibinfo {author} {\bibfnamefont {Z.~L.}\ \bibnamefont {Horváth}},
  \ and\ \bibinfo {author} {\bibfnamefont {M.~A.}\ \bibnamefont {Porras}},\
  }\href {\doibase 10.1088/2040-8978/17/6/065612} {\bibfield  {journal}
  {\bibinfo  {journal} {Journal of Optics}\ }\textbf {\bibinfo {volume} {17}},\
  \bibinfo {pages} {065612} (\bibinfo {year} {2015}{\natexlab{a}})}\BibitemShut
  {NoStop}%
\bibitem [{\citenamefont {Porras}(2009)}]{Porras2009OL}%
  \BibitemOpen
  \bibfield  {author} {\bibinfo {author} {\bibfnamefont {M.~A.}\ \bibnamefont
  {Porras}},\ }\href {\doibase 10.1364/OL.34.001546} {\bibfield  {journal}
  {\bibinfo  {journal} {Opt. Lett.}\ }\textbf {\bibinfo {volume} {34}},\
  \bibinfo {pages} {1546} (\bibinfo {year} {2009})}\BibitemShut {NoStop}%
\bibitem [{\citenamefont {Hoff}\ \emph
  {et~al.}(2017{\natexlab{a}})\citenamefont {Hoff}, \citenamefont {Kruger},
  \citenamefont {Maisenbacher}, \citenamefont {Sayler}, \citenamefont
  {Paulus},\ and\ \citenamefont {Hommelhoff}}]{Hoff2017NPhys}%
  \BibitemOpen
  \bibfield  {author} {\bibinfo {author} {\bibfnamefont {D.}~\bibnamefont
  {Hoff}}, \bibinfo {author} {\bibfnamefont {M.}~\bibnamefont {Kruger}},
  \bibinfo {author} {\bibfnamefont {L.}~\bibnamefont {Maisenbacher}}, \bibinfo
  {author} {\bibfnamefont {A.~M.}\ \bibnamefont {Sayler}}, \bibinfo {author}
  {\bibfnamefont {G.~G.}\ \bibnamefont {Paulus}}, \ and\ \bibinfo {author}
  {\bibfnamefont {P.}~\bibnamefont {Hommelhoff}},\ }\href {\doibase
  10.1038/nphys4185} {\bibfield  {journal} {\bibinfo  {journal} {Nature Phys.}\
  }\textbf {\bibinfo {volume} {13}},\ \bibinfo {pages} {947} (\bibinfo {year}
  {2017}{\natexlab{a}})}\BibitemShut {NoStop}%
\bibitem [{\citenamefont {Lindner}\ \emph {et~al.}(2004)\citenamefont
  {Lindner}, \citenamefont {Paulus}, \citenamefont {Walther}, \citenamefont
  {Baltuska}, \citenamefont {Goulielmakis}, \citenamefont {Lezius},\ and\
  \citenamefont {Krausz}}]{Lindner2004PRL}%
  \BibitemOpen
  \bibfield  {author} {\bibinfo {author} {\bibfnamefont {F.}~\bibnamefont
  {Lindner}}, \bibinfo {author} {\bibfnamefont {G.~G.}\ \bibnamefont {Paulus}},
  \bibinfo {author} {\bibfnamefont {H.}~\bibnamefont {Walther}}, \bibinfo
  {author} {\bibfnamefont {A.}~\bibnamefont {Baltuska}}, \bibinfo {author}
  {\bibfnamefont {E.}~\bibnamefont {Goulielmakis}}, \bibinfo {author}
  {\bibfnamefont {M.}~\bibnamefont {Lezius}}, \ and\ \bibinfo {author}
  {\bibfnamefont {F.}~\bibnamefont {Krausz}},\ }\href {\doibase
  10.1103/PhysRevLett.92.113001} {\bibfield  {journal} {\bibinfo  {journal}
  {Phys. Rev. Lett.}\ }\textbf {\bibinfo {volume} {92}},\ \bibinfo {pages}
  {113001} (\bibinfo {year} {2004})}\BibitemShut {NoStop}%
\bibitem [{\citenamefont {Wang}\ \emph {et~al.}(2007)\citenamefont {Wang},
  \citenamefont {Zeng}, \citenamefont {Li},\ and\ \citenamefont
  {Xu}}]{Wang2007OL}%
  \BibitemOpen
  \bibfield  {author} {\bibinfo {author} {\bibfnamefont {Z.}~\bibnamefont
  {Wang}}, \bibinfo {author} {\bibfnamefont {Z.}~\bibnamefont {Zeng}}, \bibinfo
  {author} {\bibfnamefont {R.}~\bibnamefont {Li}}, \ and\ \bibinfo {author}
  {\bibfnamefont {Z.}~\bibnamefont {Xu}},\ }\href
  {http://col.osa.org/abstract.cfm?URI=col-5-101-S183} {\bibfield  {journal}
  {\bibinfo  {journal} {Chin. Opt. Lett.}\ }\textbf {\bibinfo {volume} {5}},\
  \bibinfo {pages} {S183} (\bibinfo {year} {2007})}\BibitemShut {NoStop}%
\bibitem [{\citenamefont {Major}\ \emph
  {et~al.}(2015{\natexlab{b}})\citenamefont {Major}, \citenamefont {Nemes},
  \citenamefont {Porras}, \citenamefont {Horv\'{a}th},\ and\ \citenamefont
  {Kov\'{a}cs}}]{Major2015AO}%
  \BibitemOpen
  \bibfield  {author} {\bibinfo {author} {\bibfnamefont {B.}~\bibnamefont
  {Major}}, \bibinfo {author} {\bibfnamefont {D.}~\bibnamefont {Nemes}},
  \bibinfo {author} {\bibfnamefont {M.~A.}\ \bibnamefont {Porras}}, \bibinfo
  {author} {\bibfnamefont {Z.~L.}\ \bibnamefont {Horv\'{a}th}}, \ and\ \bibinfo
  {author} {\bibfnamefont {A.~P.}\ \bibnamefont {Kov\'{a}cs}},\ }\href
  {\doibase 10.1364/AO.54.010717} {\bibfield  {journal} {\bibinfo  {journal}
  {Appl. Opt.}\ }\textbf {\bibinfo {volume} {54}},\ \bibinfo {pages} {10717}
  (\bibinfo {year} {2015}{\natexlab{b}})}\BibitemShut {NoStop}%
\bibitem [{\citenamefont {Zapata-Rodriguez}\ and\ \citenamefont
  {Porras}(2008)}]{Zapata-Rodriguez2008OE}%
  \BibitemOpen
  \bibfield  {author} {\bibinfo {author} {\bibfnamefont {C.~J.}\ \bibnamefont
  {Zapata-Rodriguez}}\ and\ \bibinfo {author} {\bibfnamefont {M.~A.}\
  \bibnamefont {Porras}},\ }\href {\doibase 10.1364/OE.16.022090} {\bibfield
  {journal} {\bibinfo  {journal} {Optics Express}\ }\textbf {\bibinfo {volume}
  {16}},\ \bibinfo {pages} {22090} (\bibinfo {year} {2008})}\BibitemShut
  {NoStop}%
\bibitem [{\citenamefont {Porras}\ and\ \citenamefont
  {Dombi}(2009)}]{Porras2009OE}%
  \BibitemOpen
  \bibfield  {author} {\bibinfo {author} {\bibfnamefont {M.~A.}\ \bibnamefont
  {Porras}}\ and\ \bibinfo {author} {\bibfnamefont {P.}~\bibnamefont {Dombi}},\
  }\href {\doibase 10.1364/OE.17.019424} {\bibfield  {journal} {\bibinfo
  {journal} {Opt. Express}\ }\textbf {\bibinfo {volume} {17}},\ \bibinfo
  {pages} {19424} (\bibinfo {year} {2009})}\BibitemShut {NoStop}%
\bibitem [{\citenamefont {Porras}\ \emph
  {et~al.}(2012{\natexlab{a}})\citenamefont {Porras}, \citenamefont {Horvath},\
  and\ \citenamefont {Major}}]{Porras2012APB}%
  \BibitemOpen
  \bibfield  {author} {\bibinfo {author} {\bibfnamefont {M.~A.}\ \bibnamefont
  {Porras}}, \bibinfo {author} {\bibfnamefont {Z.~L.}\ \bibnamefont {Horvath}},
  \ and\ \bibinfo {author} {\bibfnamefont {B.}~\bibnamefont {Major}},\ }\href
  {\doibase 10.1007/s00340-012-5073-y} {\bibfield  {journal} {\bibinfo
  {journal} {Appl. Phys. B: Lasers Opt.}\ }\textbf {\bibinfo {volume} {108}},\
  \bibinfo {pages} {521} (\bibinfo {year} {2012}{\natexlab{a}})}\BibitemShut
  {NoStop}%
\bibitem [{\citenamefont {Porras}\ \emph
  {et~al.}(2012{\natexlab{b}})\citenamefont {Porras}, \citenamefont {Major},\
  and\ \citenamefont {Horvath}}]{Porras2012JOSAB}%
  \BibitemOpen
  \bibfield  {author} {\bibinfo {author} {\bibfnamefont {M.~A.}\ \bibnamefont
  {Porras}}, \bibinfo {author} {\bibfnamefont {B.}~\bibnamefont {Major}}, \
  and\ \bibinfo {author} {\bibfnamefont {Z.~L.}\ \bibnamefont {Horvath}},\
  }\href {\doibase 10.1364/JOSAB.29.003271} {\bibfield  {journal} {\bibinfo
  {journal} {J. Opt. Soc. Am. B}\ }\textbf {\bibinfo {volume} {29}},\ \bibinfo
  {pages} {3271} (\bibinfo {year} {2012}{\natexlab{b}})}\BibitemShut {NoStop}%
\bibitem [{\citenamefont {Major}(2016)}]{MajorPhD}%
  \BibitemOpen
  \bibfield  {author} {\bibinfo {author} {\bibfnamefont {B.}~\bibnamefont
  {Major}},\ }\emph {\bibinfo {title} {Phase and polarization changes of pulsed
  {Gaussian} beams during focusing and propagation}},\ \href {\doibase
  10.14232/phd.3115} {Ph.D. thesis},\ \bibinfo  {school} {Doctoral School of
  Physics, University of Szeged} (\bibinfo {year} {2016})\BibitemShut {NoStop}%
\bibitem [{\citenamefont {Holgado}\ \emph {et~al.}(2017)\citenamefont
  {Holgado}, \citenamefont {Hern\'andez-Garc\'{\i}a}, \citenamefont {Alonso},
  \citenamefont {Miranda}, \citenamefont {Silva}, \citenamefont {Varela},
  \citenamefont {Hern\'andez-Toro}, \citenamefont {Plaja}, \citenamefont
  {Crespo},\ and\ \citenamefont {Sola}}]{Holgado2017PRA}%
  \BibitemOpen
  \bibfield  {author} {\bibinfo {author} {\bibfnamefont {W.}~\bibnamefont
  {Holgado}}, \bibinfo {author} {\bibfnamefont {C.}~\bibnamefont
  {Hern\'andez-Garc\'{\i}a}}, \bibinfo {author} {\bibfnamefont
  {B.}~\bibnamefont {Alonso}}, \bibinfo {author} {\bibfnamefont
  {M.}~\bibnamefont {Miranda}}, \bibinfo {author} {\bibfnamefont
  {F.}~\bibnamefont {Silva}}, \bibinfo {author} {\bibfnamefont
  {O.}~\bibnamefont {Varela}}, \bibinfo {author} {\bibfnamefont
  {J.}~\bibnamefont {Hern\'andez-Toro}}, \bibinfo {author} {\bibfnamefont
  {L.}~\bibnamefont {Plaja}}, \bibinfo {author} {\bibfnamefont
  {H.}~\bibnamefont {Crespo}}, \ and\ \bibinfo {author} {\bibfnamefont {I.~J.}\
  \bibnamefont {Sola}},\ }\href {\doibase 10.1103/PhysRevA.95.063823}
  {\bibfield  {journal} {\bibinfo  {journal} {Phys. Rev. A}\ }\textbf {\bibinfo
  {volume} {95}},\ \bibinfo {pages} {063823} (\bibinfo {year}
  {2017})}\BibitemShut {NoStop}%
\bibitem [{\citenamefont {Sali\`eres}\ \emph {et~al.}(1998)\citenamefont
  {Sali\`eres}, \citenamefont {Antoine}, \citenamefont {de~Bohan},\ and\
  \citenamefont {Lewenstein}}]{Salieres1998PRL}%
  \BibitemOpen
  \bibfield  {author} {\bibinfo {author} {\bibfnamefont {P.}~\bibnamefont
  {Sali\`eres}}, \bibinfo {author} {\bibfnamefont {P.}~\bibnamefont {Antoine}},
  \bibinfo {author} {\bibfnamefont {A.}~\bibnamefont {de~Bohan}}, \ and\
  \bibinfo {author} {\bibfnamefont {M.}~\bibnamefont {Lewenstein}},\ }\href
  {\doibase 10.1103/PhysRevLett.81.5544} {\bibfield  {journal} {\bibinfo
  {journal} {Phys. Rev. Lett.}\ }\textbf {\bibinfo {volume} {81}},\ \bibinfo
  {pages} {5544} (\bibinfo {year} {1998})}\BibitemShut {NoStop}%
\bibitem [{\citenamefont {Chang}\ \emph {et~al.}(1998)\citenamefont {Chang},
  \citenamefont {Rundquist}, \citenamefont {Wang}, \citenamefont {Christov},
  \citenamefont {Kapteyn},\ and\ \citenamefont {Murnane}}]{Chang1998PRA}%
  \BibitemOpen
  \bibfield  {author} {\bibinfo {author} {\bibfnamefont {Z.}~\bibnamefont
  {Chang}}, \bibinfo {author} {\bibfnamefont {A.}~\bibnamefont {Rundquist}},
  \bibinfo {author} {\bibfnamefont {H.}~\bibnamefont {Wang}}, \bibinfo {author}
  {\bibfnamefont {I.}~\bibnamefont {Christov}}, \bibinfo {author}
  {\bibfnamefont {H.~C.}\ \bibnamefont {Kapteyn}}, \ and\ \bibinfo {author}
  {\bibfnamefont {M.~M.}\ \bibnamefont {Murnane}},\ }\href {\doibase
  10.1103/PhysRevA.58.R30} {\bibfield  {journal} {\bibinfo  {journal} {Phys.
  Rev. A: At., Mol., Opt. Phys.}\ }\textbf {\bibinfo {volume} {58}},\ \bibinfo
  {pages} {R30} (\bibinfo {year} {1998})}\BibitemShut {NoStop}%
\bibitem [{\citenamefont {Lee}\ \emph {et~al.}(2001)\citenamefont {Lee},
  \citenamefont {Kim}, \citenamefont {Hong},\ and\ \citenamefont
  {Nam}}]{Lee2001PRL}%
  \BibitemOpen
  \bibfield  {author} {\bibinfo {author} {\bibfnamefont {D.~G.}\ \bibnamefont
  {Lee}}, \bibinfo {author} {\bibfnamefont {J.-H.}\ \bibnamefont {Kim}},
  \bibinfo {author} {\bibfnamefont {K.-H.}\ \bibnamefont {Hong}}, \ and\
  \bibinfo {author} {\bibfnamefont {C.~H.}\ \bibnamefont {Nam}},\ }\href
  {\doibase 10.1103/PhysRevLett.87.243902} {\bibfield  {journal} {\bibinfo
  {journal} {Phys. Rev. Lett.}\ }\textbf {\bibinfo {volume} {87}},\ \bibinfo
  {pages} {243902} (\bibinfo {year} {2001})}\BibitemShut {NoStop}%
\bibitem [{\citenamefont {Mauritsson}\ \emph {et~al.}(2004)\citenamefont
  {Mauritsson}, \citenamefont {Johnsson}, \citenamefont {L\'opez-Martens},
  \citenamefont {Varj\'u}, \citenamefont {Kornelis}, \citenamefont {Biegert},
  \citenamefont {Keller}, \citenamefont {Gaarde}, \citenamefont {Schafer},\
  and\ \citenamefont {L'Huillier}}]{Mauritssom2004PRA}%
  \BibitemOpen
  \bibfield  {author} {\bibinfo {author} {\bibfnamefont {J.}~\bibnamefont
  {Mauritsson}}, \bibinfo {author} {\bibfnamefont {P.}~\bibnamefont
  {Johnsson}}, \bibinfo {author} {\bibfnamefont {R.}~\bibnamefont
  {L\'opez-Martens}}, \bibinfo {author} {\bibfnamefont {K.}~\bibnamefont
  {Varj\'u}}, \bibinfo {author} {\bibfnamefont {W.}~\bibnamefont {Kornelis}},
  \bibinfo {author} {\bibfnamefont {J.}~\bibnamefont {Biegert}}, \bibinfo
  {author} {\bibfnamefont {U.}~\bibnamefont {Keller}}, \bibinfo {author}
  {\bibfnamefont {M.~B.}\ \bibnamefont {Gaarde}}, \bibinfo {author}
  {\bibfnamefont {K.~J.}\ \bibnamefont {Schafer}}, \ and\ \bibinfo {author}
  {\bibfnamefont {A.}~\bibnamefont {L'Huillier}},\ }\href {\doibase
  10.1103/PhysRevA.70.021801} {\bibfield  {journal} {\bibinfo  {journal} {Phys.
  Rev. A}\ }\textbf {\bibinfo {volume} {70}},\ \bibinfo {pages} {021801}
  (\bibinfo {year} {2004})}\BibitemShut {NoStop}%
\bibitem [{\citenamefont {Holgado}\ \emph {et~al.}(2016)\citenamefont
  {Holgado}, \citenamefont {Hern\'andez-Garc\'{\i}a}, \citenamefont {Alonso},
  \citenamefont {Miranda}, \citenamefont {Silva}, \citenamefont {Plaja},
  \citenamefont {Crespo},\ and\ \citenamefont {Sola}}]{Holgado2016PRA}%
  \BibitemOpen
  \bibfield  {author} {\bibinfo {author} {\bibfnamefont {W.}~\bibnamefont
  {Holgado}}, \bibinfo {author} {\bibfnamefont {C.}~\bibnamefont
  {Hern\'andez-Garc\'{\i}a}}, \bibinfo {author} {\bibfnamefont
  {B.}~\bibnamefont {Alonso}}, \bibinfo {author} {\bibfnamefont
  {M.}~\bibnamefont {Miranda}}, \bibinfo {author} {\bibfnamefont
  {F.}~\bibnamefont {Silva}}, \bibinfo {author} {\bibfnamefont
  {L.}~\bibnamefont {Plaja}}, \bibinfo {author} {\bibfnamefont
  {H.}~\bibnamefont {Crespo}}, \ and\ \bibinfo {author} {\bibfnamefont {I.~J.}\
  \bibnamefont {Sola}},\ }\href {\doibase 10.1103/PhysRevA.93.013816}
  {\bibfield  {journal} {\bibinfo  {journal} {Phys. Rev. A: At., Mol., Opt.
  Phys.}\ }\textbf {\bibinfo {volume} {93}},\ \bibinfo {pages} {013816}
  (\bibinfo {year} {2016})}\BibitemShut {NoStop}%
\bibitem [{\citenamefont {Goswami}\ \emph {et~al.}(2009)\citenamefont
  {Goswami}, \citenamefont {Kumar}, \citenamefont {Dutta},\ and\ \citenamefont
  {Goswami}}]{Goswami2009CP}%
  \BibitemOpen
  \bibfield  {author} {\bibinfo {author} {\bibfnamefont {T.}~\bibnamefont
  {Goswami}}, \bibinfo {author} {\bibfnamefont {S.~K.}\ \bibnamefont {Kumar}},
  \bibinfo {author} {\bibfnamefont {A.}~\bibnamefont {Dutta}}, \ and\ \bibinfo
  {author} {\bibfnamefont {D.}~\bibnamefont {Goswami}},\ }\href {\doibase
  https://doi.org/10.1016/j.chemphys.2009.04.009} {\bibfield  {journal}
  {\bibinfo  {journal} {Chemical Physics}\ }\textbf {\bibinfo {volume} {360}},\
  \bibinfo {pages} {47 } (\bibinfo {year} {2009})}\BibitemShut {NoStop}%
\bibitem [{\citenamefont {Vidal}\ \emph {et~al.}(2014)\citenamefont {Vidal},
  \citenamefont {Degert}, \citenamefont {Tondusson}, \citenamefont {Freysz},\
  and\ \citenamefont {Oberl\'{e}}}]{Vidal2014JOSAB}%
  \BibitemOpen
  \bibfield  {author} {\bibinfo {author} {\bibfnamefont {S.}~\bibnamefont
  {Vidal}}, \bibinfo {author} {\bibfnamefont {J.}~\bibnamefont {Degert}},
  \bibinfo {author} {\bibfnamefont {M.}~\bibnamefont {Tondusson}}, \bibinfo
  {author} {\bibfnamefont {E.}~\bibnamefont {Freysz}}, \ and\ \bibinfo {author}
  {\bibfnamefont {J.}~\bibnamefont {Oberl\'{e}}},\ }\href {\doibase
  10.1364/JOSAB.31.000149} {\bibfield  {journal} {\bibinfo  {journal} {J. Opt.
  Soc. Am. B}\ }\textbf {\bibinfo {volume} {31}},\ \bibinfo {pages} {149}
  (\bibinfo {year} {2014})}\BibitemShut {NoStop}%
\bibitem [{\citenamefont {Karimi}\ \emph {et~al.}(2013)\citenamefont {Karimi},
  \citenamefont {Altucci}, \citenamefont {Tosa}, \citenamefont {Velotta},\ and\
  \citenamefont {Marrucci}}]{Karimi2013OE}%
  \BibitemOpen
  \bibfield  {author} {\bibinfo {author} {\bibfnamefont {E.}~\bibnamefont
  {Karimi}}, \bibinfo {author} {\bibfnamefont {C.}~\bibnamefont {Altucci}},
  \bibinfo {author} {\bibfnamefont {V.}~\bibnamefont {Tosa}}, \bibinfo {author}
  {\bibfnamefont {R.}~\bibnamefont {Velotta}}, \ and\ \bibinfo {author}
  {\bibfnamefont {L.}~\bibnamefont {Marrucci}},\ }\href {\doibase
  10.1364/OE.21.024991} {\bibfield  {journal} {\bibinfo  {journal} {Opt.
  Express}\ }\textbf {\bibinfo {volume} {21}},\ \bibinfo {pages} {24991}
  (\bibinfo {year} {2013})}\BibitemShut {NoStop}%
\bibitem [{\citenamefont {Hoff}\ \emph
  {et~al.}(2017{\natexlab{b}})\citenamefont {Hoff}, \citenamefont {Krüger},
  \citenamefont {Maisenbacher}, \citenamefont {Paulus}, \citenamefont
  {Hommelhoff},\ and\ \citenamefont {Sayler}}]{Hoff2017JOP}%
  \BibitemOpen
  \bibfield  {author} {\bibinfo {author} {\bibfnamefont {D.}~\bibnamefont
  {Hoff}}, \bibinfo {author} {\bibfnamefont {M.}~\bibnamefont {Krüger}},
  \bibinfo {author} {\bibfnamefont {L.}~\bibnamefont {Maisenbacher}}, \bibinfo
  {author} {\bibfnamefont {G.~G.}\ \bibnamefont {Paulus}}, \bibinfo {author}
  {\bibfnamefont {P.}~\bibnamefont {Hommelhoff}}, \ and\ \bibinfo {author}
  {\bibfnamefont {A.~M.}\ \bibnamefont {Sayler}},\ }\href {\doibase
  10.1088/2040-8986/aa9247} {\bibfield  {journal} {\bibinfo  {journal} {Journal
  of Optics}\ }\textbf {\bibinfo {volume} {19}},\ \bibinfo {pages} {124007}
  (\bibinfo {year} {2017}{\natexlab{b}})}\BibitemShut {NoStop}%
\bibitem [{\citenamefont {Porras}(2002)}]{Porras2002PRE}%
  \BibitemOpen
  \bibfield  {author} {\bibinfo {author} {\bibfnamefont {M.~A.}\ \bibnamefont
  {Porras}},\ }\href {\doibase 10.1103/PhysRevE.65.026606} {\bibfield
  {journal} {\bibinfo  {journal} {Phys. Rev. E}\ }\textbf {\bibinfo {volume}
  {65}},\ \bibinfo {pages} {026606} (\bibinfo {year} {2002})}\BibitemShut
  {NoStop}%
\bibitem [{\citenamefont {Bor}\ and\ \citenamefont
  {Horv\'{a}th}(1992)}]{Bor1992OC}%
  \BibitemOpen
  \bibfield  {author} {\bibinfo {author} {\bibfnamefont {Z.}~\bibnamefont
  {Bor}}\ and\ \bibinfo {author} {\bibfnamefont {Z.~L.}\ \bibnamefont
  {Horv\'{a}th}},\ }\href {\doibase
  http://dx.doi.org/10.1016/0030-4018(92)90022-J} {\bibfield  {journal}
  {\bibinfo  {journal} {Optics Communications}\ }\textbf {\bibinfo {volume}
  {94}},\ \bibinfo {pages} {249} (\bibinfo {year} {1992})}\BibitemShut
  {NoStop}%
\bibitem [{\citenamefont {Gauthier}\ \emph {et~al.}(2017)\citenamefont
  {Gauthier}, \citenamefont {Ribic}, \citenamefont {Adhikary}, \citenamefont
  {Camper}, \citenamefont {Chappuis}, \citenamefont {Cucini}, \citenamefont
  {DiMauro}, \citenamefont {Dovillaire}, \citenamefont {Frassetto},
  \citenamefont {G{\'e}neaux}, \citenamefont {Miotti}, \citenamefont {Poletto},
  \citenamefont {Ressel}, \citenamefont {Spezzani}, \citenamefont {Stupar},
  \citenamefont {Ruchon},\ and\ \citenamefont {De~Ninno}}]{Gauthier2017Nature}%
  \BibitemOpen
  \bibfield  {author} {\bibinfo {author} {\bibfnamefont {D.}~\bibnamefont
  {Gauthier}}, \bibinfo {author} {\bibfnamefont {P.~R.}\ \bibnamefont {Ribic}},
  \bibinfo {author} {\bibfnamefont {G.}~\bibnamefont {Adhikary}}, \bibinfo
  {author} {\bibfnamefont {A.}~\bibnamefont {Camper}}, \bibinfo {author}
  {\bibfnamefont {C.}~\bibnamefont {Chappuis}}, \bibinfo {author}
  {\bibfnamefont {R.}~\bibnamefont {Cucini}}, \bibinfo {author} {\bibfnamefont
  {L.~F.}\ \bibnamefont {DiMauro}}, \bibinfo {author} {\bibfnamefont
  {G.}~\bibnamefont {Dovillaire}}, \bibinfo {author} {\bibfnamefont
  {F.}~\bibnamefont {Frassetto}}, \bibinfo {author} {\bibfnamefont
  {R.}~\bibnamefont {G{\'e}neaux}}, \bibinfo {author} {\bibfnamefont
  {P.}~\bibnamefont {Miotti}}, \bibinfo {author} {\bibfnamefont
  {L.}~\bibnamefont {Poletto}}, \bibinfo {author} {\bibfnamefont
  {B.}~\bibnamefont {Ressel}}, \bibinfo {author} {\bibfnamefont
  {C.}~\bibnamefont {Spezzani}}, \bibinfo {author} {\bibfnamefont
  {M.}~\bibnamefont {Stupar}}, \bibinfo {author} {\bibfnamefont
  {T.}~\bibnamefont {Ruchon}}, \ and\ \bibinfo {author} {\bibfnamefont
  {G.}~\bibnamefont {De~Ninno}},\ }\href {\doibase 10.1038/ncomms14971}
  {\bibfield  {journal} {\bibinfo  {journal} {Nature Communications}\ }\textbf
  {\bibinfo {volume} {8}},\ \bibinfo {pages} {14971} (\bibinfo {year}
  {2017})}\BibitemShut {NoStop}%
\bibitem [{\citenamefont {Biss}\ and\ \citenamefont
  {Brown}(2003)}]{Biss2003OL}%
  \BibitemOpen
  \bibfield  {author} {\bibinfo {author} {\bibfnamefont {D.~P.}\ \bibnamefont
  {Biss}}\ and\ \bibinfo {author} {\bibfnamefont {T.~G.}\ \bibnamefont
  {Brown}},\ }\href {\doibase 10.1364/OL.28.000923} {\bibfield  {journal}
  {\bibinfo  {journal} {Optics Letters}\ }\textbf {\bibinfo {volume} {28}},\
  \bibinfo {pages} {923} (\bibinfo {year} {2003})}\BibitemShut {NoStop}%
\end{thebibliography}%

\end{document}